\documentclass[11pt]{article}

\usepackage{mathrsfs}
\usepackage{dsfont}
\usepackage{amssymb}
\usepackage{float}
\usepackage{enumitem}
\usepackage{tikz}
\usepackage{chet}

\usetikzlibrary{decorations.markings,arrows,calc}

\tikzset{>=latex,
         ->/.style={decoration={markings,mark=at position 1 with
           {\arrow[scale=1.5]{>}}}, postaction={decorate}}}
\tikzset{->-/.style={decoration={markings,mark=at position 0.5 with
           {\arrow[scale=1.5]{>}}},postaction={decorate}}}
\tikzset{-<-/.style={decoration={markings,mark=at position 0.5 with
           {\arrow[scale=1.5]{<}}},postaction={decorate}}}
\tikzset{cross/.style={path picture={
      \draw[black]
            (path picture bounding box.south east) --
            (path picture bounding box.north west)
            (path picture bounding box.south west) --
            (path picture bounding box.north east);}}}

\newcommand{\vev}[1]{\langle {#1}\rangle}

\newcommand{\CO}{\mathcal{O}}
\newcommand{\hpartial}{\hat{\partial}}

\DeclareMathOperator{\tr}{tr}
\DeclareMathOperator{\Ima}{Im}

\date{November 2012}

\preprint{CERN-PH-TH/2012-297\\
SU-ITP-12/38\\
UCSD-PTH-12-10}

\title{Limit Cycles and Conformal Invariance}

\author{Jean-Fran\c{c}ois Fortin,$^{\ast,\dag,\$,}$\email{jean-francois.fortin@cern.ch}
Benjam\'\i{}n Grinstein$^{\$,}$\email{bgrinstein@ucsd.edu}
and Andreas Stergiou$^{\$,}$\email{stergiou@physics.ucsd.edu}}

\affiliation{$^{\ast}$Theory Division, Department of Physics, CERN, CH-1211 Geneva 23, Switzerland\\
$^\dag$Stanford Institute for Theoretical Physics, Department of Physics, Stanford University, Stanford, CA 94305, USA\\
$^{\$}$Department of Physics, University of California, San Diego, La Jolla, CA 92093, USA}

\abstract{There is a widely held belief that conformal field theories
(CFTs) require zero beta functions.  Nevertheless, the work of Jack and
Osborn implies that the beta functions are not actually the quantites that
decide conformality, but until recently no such behavior had been
exhibited.  Our recent work has led to the discovery of CFTs with nonzero
beta functions, more precisely CFTs that live on recurrent trajectories,
e.g., limit cycles, of the beta-function vector field. To demonstrate this
we study the $S$ function of Jack and Osborn.  We use Weyl consistency
conditions to show that it vanishes at fixed points and agrees with the
generator $Q$ of limit cycles on them.  Moreover, we compute $S$ to third
order in perturbation theory, and explicitly verify that it agrees with our
previous determinations of $Q$. A byproduct of our analysis is that, in
perturbation theory, unitarity and scale invariance imply conformal
invariance in four-dimensional quantum field theories.  Finally, we study
some properties of these new, ``cyclic'' CFTs, and point out that the
$a$-theorem still governs the asymptotic behavior of renormalization-group
flows.}

\begin{document}

\maketitle

\newsec{Overview in lieu of Introduction}
Two recent reported results can potentially greatly enrich our
understanding of quantum field theory (QFT). On the one hand, Komargodski
and Schwimmer (KS) \cite{Komargodski:2011vj}, following earlier work by
Cappelli, D'Appollonio, Guida and Magnoli (CDGM) \cite{Cappelli:2000dv,
Cappelli:2001pz}, have delineated a nonperturbative proof of an inequality
satisfied when a four-dimensional QFT flows between two fixed points of the
renormalization group (RG). On the other hand, we have discovered closed RG
trajectories\foot{Meaning closed flow-lines of the familiar dim-reg
beta-function vector field, in conventions where the anomalous-dimension
matrix is symmetric.  For a word on conventions and their effects on RG
functions see Appendix~\ref{app:Bfunction}.} in theories in $d=4-\epsilon$
\cite{Fortin:2011ks, Fortin:2011bm, Fortin:2012ic} and $d=4$
\cite{Fortin:2012cq} spacetime dimensions, in a regime where perturbation
theory is applicable.  While the former result can impose restrictions on
the possible realizations of long distance (IR) phases of QFTs, the latter
exhibits explicitly a novel feature of QFTs.  A question naturally arises
as to whether these results are compatible.

In this work we will show perturbatively that unitary, interacting,
scale-invariant cycles\foot{More precisely ``limit recursive flows'' of the
dim-reg beta-function vector field. In what follows we refer to both limit
cycles and limiting ergodic behavior simply as ``cycles.''} in $d=4$
correspond to conformal field theories (CFTs), that is, theories with
invariance under the full conformal group, not just Poincar\'{e} plus
dilatations.  This follows from the work of Jack and Osborn
(JO)~\cite{Jack:1990eb}. Compatibility of this type of cycles\footnote{The
condition for scale invariance, $\mu\,dg^i/d\mu=Q^i_{\phantom{i}\!j}g^j$,
$Q^T=-Q=\text{constant}$ \cite{Polchinski:1987dy}, gives recursive flows
\cite{Fortin:2011sz}. Our study of cycles here is concerned with this type
of closed trajectories, given by a rotation of the coupling constants by a
compact Abelian group generated by $Q$. Whether recursive flows that are
not of this type exist is an open question.} with the aforementioned
inequality is then not surprising since the inequality still compares a
quantity defined on CFTs, be it a CFT at an endpoint of an RG flow or a CFT
corresponding to a limit cycle of the RG flow.

To be clear, the cycles we discuss in this work are not associated with
unitary theories that are scale but not conformally
invariant~\cite{Fortin:2011sz, Fortin:2011bm}. In fact, in this paper we
prove that limit cycles associated with scale but not conformally invariant
unitary theories do not exist in perturbation theory. As we will see, on
cycles of the dim-reg beta function discussed here the theory is fully
conformal.  This includes the examples of~\cite{Fortin:2011ks,
Fortin:2012ic}, which were incorrectly interpreted there as examples of
theories with scale but without conformal invariance. We should point out
that, conformal theories defined on cycles of the dim-reg beta function have
the same properties as conformal theories defined at traditional fixed
points.

Luty, Polchinski and Rattazzi (LPR) \cite{Luty:2012ww} argued that limit
cycles cannot exist in $d=4$ unitary QFT, and hence that scale without
conformal invariance is excluded.  As we shall see, limit cycles do occur,
but QFTs on them are fully conformal, not just scale-invariant.  LPR have
informed us that their manuscript is being replaced with one that contains
a corrected version of their argument, with their conclusion regarding the
absence of scale without conformal invariance unchanged.

The work of KS is not sensitive to the presence of cycles.  Indeed, KS
assume the existence of a flow from a short distance (UV) CFT to an IR CFT,
and argue that the coefficient $a$ of the Euler density in the curved-space
trace anomaly,
\eqn{T^\mu_{\phantom{\mu}\!\mu}=\text{operator terms}+c(\text{Weyl tensor
squared})-a(\text{Euler density}),}
is larger at the UV than the IR fixed point: $a_\text{UV}> a_\text{IR}$.
This, then, is a proof of the ``weak version'' of the $c$-theorem.  The KS
argument incorporates putative flows from a fixed point or cycle to another
fixed point or cycle, since in both cases the theories encountered are
CFTs.

In $d=2$ a stronger result holds: there exists a quantity $c$, local in the
RG scale, that is monotonically decreasing along any RG flow
\cite{Zamolodchikov:1986gt}.  This is referred to as the ``strong version''
of the $c$-theorem, and it was first argued to also be true in $d=4$ by
Cardy \cite{Cardy:1988cwa}.  A proof was later found by JO (see also
\cite{Osborn:1991gm}), albeit only in perturbation theory.  Away from fixed
points the quantity that plays the role of $c$ in the arguments of JO is
not exactly equal to $a$ (the coefficient of the Euler density in the
curved-space trace anomaly).  However, it agrees with $a$ at
endpoints/limit cycles of the RG trajectories. This is in agreement with
the result of KS that the weak version of the $c$-theorem is valid for $a$.
In this paper we extend the perturbative proof of JO to include RG cycles.

Of course it is well-known that $a$ may {\it increase} away from trivial UV
fixed points: for example, for pure Yang--Mills (YM) theory with beta
function $\beta^g=-\beta_0g^3/16\pi^2-\beta_1g^5/(16\pi^2)^2-\cdots$ one
has \cite{Jack:1990eb}
\eqn{a={a}_0+ \frac{n_V\beta_1}{8(16\pi^2)^3}g^4+\CO(g^6).}[atwoloops]
Here $a_0$ is the free field theory (one-loop) value of $a$ and
$n_V=\dim(\text{Adj})$ is the number of vector fields.\foot{We thank K.\
Intriligator for discussions on this point.}  Nevertheless, even in this
case JO showed that there exists a quantity, $\tilde{\beta}_b$, which flows
monotonically (to all orders in perturbation theory). The quantity
$\tilde{\beta}_b$ is related to $a$, which in JO is denoted by $\beta_b$,
by
\eqn{\tilde{\beta}_b=\beta_b+\tfrac18 w\beta^g, \qquad \beta_b\equiv a.}
Here $w$ is a function of the coupling $g$, and $\beta^g=-dg/dt$ is the
beta function.  While $\tilde{\beta}_b$ and $\beta_b$ agree on fixed
points, the difference is parametrically large away from fixed points. In
Section~\ref{Sec:WeylCC} we explain this in detail.

The result of JO follows from careful inspection of how the theory responds
to Weyl rescaling.  The KS method, or an elaboration on it by
LPR~\cite{Luty:2012ww}, extensively uses Weyl rescaling and takes advantage
of the particularly simple form this takes on fixed points.  However, in
trying to extend the KS arguments to produce a proof of the strong version
of the $c$-theorem, LPR use Weyl rescaling away from fixed points. We
explain how consistency requires introducing spacetime-dependent coupling
constants and then in addition new counterterms that involve derivatives of
the couplings. We use the very rescaling in LPR to derive JO's consistency
conditions anew, of which the monotonic flow of $\tilde\beta_b$ is but one
example.

For models which display cycles the state of affairs is significantly more
complex. In all these models the kinetic terms of the Lagrangian are
invariant under a ``flavor'' symmetry group $G_F$ (that commutes with the
gauge group).  Scalar self-interactions and Yukawa couplings of scalars
with fermions break $G_F$. The dependence of counterterms on the coupling
constants characterizing these interactions is restricted by the pattern of
breaking of $G_F$. There is a well-known, simple method of accounting for
this. The coupling constants are treated as spurions, that is, as
non-dynamical fields, and allowed to transform under $G_F$ precisely so as
to render the Lagrangian invariant under these symmetry transformations.
Then, if the regulator respects the symmetry, so will the counterterms. It
follows that the entries in the trace anomaly respect the symmetry too. As
$a$ is the coefficient of the $G_F$-invariant Euler density, it is itself
$G_F$-invariant as well.  And since the flow on a cycle corresponds to a
$G_F$-transformation of the couplings, $a$ remains constant on the cycle.

This raises the following question: ``how is the monotonic flow of
$\tilde\beta_b$ consistent with the constancy of $a$?'' Actually,
$\tilde\beta_b$ is also $G_F$-invariant, and is thus also guaranteed to be
constant along the cyclic flow. The answer is found in the flow equation
for $\tilde\beta_b$ given by Osborn in \cite{Osborn:1991gm}. His equation
is a generalization, applicable to these more complex theories, of that
found by JO. This flow equation is not guaranteed to give monotonic flows,
but can and does give constancy of $\tilde\beta_b$ on cycles. We review the
work of JO concerning these more complex theories in Section
\ref{Sec:DimThree}, and show that a quantity $\widetilde B_b$ decreases
monotonically along RG flows, at least in perturbation theory, and agrees
with $\beta_b$ on fixed points and cycles.  This is a result essentially
contained in \cite{Jack:1990eb, Osborn:1991gm}, although it is not
explicitly mentioned there.

To obtain this result an understanding of the modifications to the
trace-anomaly equation in theories with cycles is required. It is a
little-known fact that in theories with many fermions and scalars there
generically appears a term in the trace anomaly of the form of the
divergence of a current.  The current generates transformations that
correspond to a particular element $S$ of the Lie algebra of $G_F$, that is
a function of the coupling constants. JO showed by direct computation that
$S$ vanishes to three loops if the field content of the QFT consists solely
of scalars, and to two loops if of scalars and Dirac spinors. On the other
hand, the element $Q$ of the Lie algebra of $G_F$ that generates the flow
along the cycle is found in our computations to arise at three-loop order
in gauge theories that include both scalars and
spinors~\cite{Fortin:2012cq}. Could it be, then, that $S$ is non-zero at
three loops in these theories? And if so, what is the relation between $Q$
and $S$? In Section~\ref{QvsS} we take on the task of computing the
lowest-order contribution to JO's $S$ for the most general four-dimensional
QFT, compare with $Q$ and demonstrate that $S$ agrees with $Q$ on cycles
and vanishes on fixed points. A corollary of this result is that scale
implies conformal invariance in relativistic unitary perturbative
four-dimensional QFTs.

That $S$ agrees with $Q$ on cycles suggests that the two terms in the
flat-space trace anomaly may cancel.  That is, the well-known
$\beta\,\partial\mathscr{L}/\partial g$ term may cancel against the
little-known divergence of the $S$-current term, since the $\beta$-term is
determined by $Q$ on cycles. This is indeed what happens: the trace of the
stress-energy tensor vanishes for unitary, scale-invariant cycles, and
hence these models display invariance under the full conformal group. In
the rest of Section~\ref{cyclicCFTs} we prove this and explore a few of its
consequences.  Armed with these results, we return to the proof of the
$c$-theorem in Section~\ref{Sec:a-theorem}. There we give a slightly
streamlined version of the LPR version of the KS argument, with care to
address the possible differences that may arise when the CFTs at the ends
of the RG flow correspond to cycles.

Let us specify here that we follow closely the notation of JO
\cite{Jack:1990eb}, with some notable exceptions. From here on, following
JO, in the anomaly equation we use $\beta_a$ and $\beta_b$ for $-c$ and $a$
respectively, although we still use the terminology $c$-theorem instead of
the more accurate $\beta_b$-theorem.  Also, we call $\lambda_{a,b,c}$
rather than $a,b,c$ the infinite counterterms that give rise to
$\beta_{a,b,c}$ (having infinite counterterms labeled by $a$ and $c$ can
certainly produce confusion with the corresponding ``beta functions'' that
appear in the Weyl anomaly and that are commonly referred to as $a$ and
$c$).  Throughout this paper RG time is defined by $t=-\ln(\mu/\mu_0)$, so
that it increases as we flow to the IR.

Many of our results are extracted from the work of JO.  So it is perhaps
necessary to remark that, besides parsing the results of JO to hopefully
make them slightly more accessible to the general reader, we have made
several novel contributions:
\begin{itemize}
  \item We have discovered where in the argument of LPR the quantity
    $\beta_b$ is replaced by $\tilde{\beta}_b$ (or, in
    more generality, by $\widetilde{B}_b$).
  \item We have extended JO's calculation of $S$ to third loop order,
    which is the leading non-vanishing contribution to $S$ in a
    Yang--Mills theory with scalars and spinors.
  \item We have shown that $S$ vanishes on fixed points and agrees with
    the generator $Q$ of limit cycles on them.
  \item We have demonstrated in perturbation theory that
     unitary, scale and Poincar\'{e}
    invariant, interacting QFTs in $d=4$ have vanishing trace of the
    stress-energy tensor and hence are invariant under the full conformal
    group.
  \item We have used the above to
  \begin{itemize}[label=$\circ$]
    \item find, using arguments of JO, a perturbative proof of an extension
      of the strong version of the $c$-theorem, i.e., that there exists a
      quantity that monotonically decreases in flows out of UV fixed points
      and cycles, and
    \item clarify that the arguments of KS apply even in the presence of
      cycles, i.e., that $(\beta_b)_\text{UV}>(\beta_b)_\text{IR}$ for
      presumed RG flows that can now originate or terminate on cycles as
      well as fixed points, valid even outside perturbation theory
      (provided the implicit assumptions in KS do not invalidate their
      result).
  \end{itemize}
\end{itemize}

\newsec{Weyl consistency conditions}
\label{Sec:WeylCC}
In this section we review the derivation of the Weyl consistency conditions
of JO. The method uses as a starting point the expressions of Weyl
invariance used by KS and by LPR. The presentation is formulated so that it
becomes clear that the assumptions in those works already implicitly lead
to the JO consistency conditions. Hence, although the derivation presented
here may seem novel, it actually follows closely JO. We have included it
here for completeness, for pedagogy and because it makes clear that neither
the results of KS nor those of LPR should be in conflict with those of JO.

Let us briefly review Osborn's argument for the consistency conditions
\cite{Osborn:1991gm}. These are analogous to the well-known Wess--Zumino
consistency conditions \cite{Wess:1971yu}. Let $\Delta_\sigma\widetilde W$
denote the action of a Weyl transformation on $\widetilde W$, the
generating functional for connected renormalized Green functions.  Because
of the Abelian nature of the Weyl group, the Weyl consistency conditions
follow:
\eqn{[\Delta_\sigma,\Delta_{\sigma^\prime}]\widetilde W=0.}
In JO the same consistency conditions are obtained by requiring finiteness
of the trace of the stress-energy tensor in curved background and with
spacetime-dependent couplings. One can also obtain the Weyl consistency
conditions based on the arguments of LPR.

LPR start from a quantum action $S_0$ which is a function of a conformally
flat metric, $\gamma_{\mu\nu}=e^{-2\tau(x)}\eta_{\mu\nu}$ and coupling
constants $g^i(\mu)$ (in $d=4-\epsilon$ regularization, with, say, minimal
subtraction (MS)). By rescaling the fields, which are dummy variables of
integration anyway, by $\phi\to(e^{\tau})^\delta\phi$, where $\delta$ is
the canonical dimension of the field (as in $\delta=(d-2)/2$ for scalars),
and using the $\mu$-independence of the bare couplings, one sees that the
$\tau$-dependence in $S_0$ arises only due to the scale dependence of
renormalized coupling constants, $g^i(e^{\tau}\mu)$.  Effectively, the
regularized generating functional $W$ satisfies
\eqn{ W[e^{-2\tau(x)}\eta_{\mu\nu},
g^i(\mu)]=  W[\eta_{\mu\nu}, g^i(e^{\tau(x)}\mu)].}[LPRinvariance]
Alternatively, Komargodski \cite{Komargodski:2011xv} argues that the
functional is made invariant under Weyl transformations by adding a
conformal compensator $\tau(x)$.  One can write
\eqn{ W[e^{-2\tau(x)}\eta_{\mu\nu},
  g^i(e^{-\tau(x)}\mu)]=  W[\eta_{\mu\nu}, g^i(\mu)],}
or, equivalently, that the left-hand side is invariant under
$\tau\to\tau+\sigma$. For finiteness it is also necessary to include in $W$
all possible counterterms that are functions of spacetime-dependent
background and coupling constants, $\gamma_{\mu\nu}(x)$ and $g^i(x)$. It is
from counterterms that do not vanish for spacetime-independent coupling
constants that the $\beta_{a,b,c}$-anomalies arise. It is convenient, in
order to keep track of curvature-dependent terms, to do this in a more
general background metric,
\twoseqn{W[e^{-2\tau(x)}\gamma_{\mu\nu}(x), g^i(\mu)]&=W[\gamma_{\mu\nu}(x),
g^i(e^{\tau(x)}\mu)],}[FULLinva]{ W[e^{-2\tau(x)}\gamma_{\mu\nu}(x),
g^i(e^{-\tau(x)}\mu)]&= W[\gamma_{\mu\nu}(x),
g^i(\mu)].}[FULLinvb][FULLinv]
At the risk of restating the trivial, let us emphasize that it is not
consistent to neglect the spacetime dependence of couplings when studying
Weyl anomalies, since the Weyl transformation involves spacetime-dependent
couplings. The counterterms associated with spacetime derivatives of these
couplings will lead to additional anomalies. Some of these may---and as we
will see, do---contribute to the dilaton/compensator scattering amplitude
even after one takes the limit of flat background and spacetime-independent
coupling constants.

The approach of LPR allows one to compute quantities associated with a
model in a curved background with spacetime-independent coupling constants
in terms of corresponding quantities for the same model but in a flat
background with, however, spacetime-dependent coupling constants. This
observation is not new. For example, in JO the same observation is used
precisely for the same purpose, namely, to compute the anomalies associated
with scale transformations using only computations in flat space.
Similarly, the approach of Komargodski allows for an explicit nonlinear
realization of scale invariance, at the price of introducing
spacetime-dependent coupling constants. In either case it is important to
realize that new counterterms are required to render the model finite, much
like counterterms involving derivatives of the metric need to be introduced
to render finite the model in a curved background. These new counterterms
must involve derivatives of the coupling constants and lead to new Weyl
anomalies. At the end of this section we study how these new anomalies
contribute to the Wess--Zumino action for the conformal compensator
$\tau(x)$ even after the couplings and the metric are taken to be
spacetime-independent. For the remainder of this section we take a closer
look at these counterterms, the anomalies they produce and the relations
between them, that is, the JO consistency conditions,  that follow from
\FULLinv.

Consider the theory in the background of an arbitrary metric
$\gamma_{\mu\nu}(x)$ and arbitrary spacetime-dependent coupling constants
$g^i(x)$ corresponding to interaction terms $g^i(x)\CO_i(x)$ in the
Lagrangian. The arbitrary spacetime dependence of the couplings allows one
to use them as sources for operators in the interaction part of the
Lagrangian, by taking functional derivatives of the generating functional
with respect to $g^i(x)$. If the quantum action is renormalized, then this
procedure automatically gives finite correlations functions for the
insertions of these operators.  Let $\widetilde W$ stand for the
renormalized generating functional. It is convenient to separate the
counterterms that are independent of quantum fields from the rest of the
action. They can be taken out of the functional integral and contribute
directly to $\widetilde W$:
\eqn{\widetilde W=W+W_{\text{c.t.}}.}
The generating functional $W$ results from performing the functional
integral over quantum fields in the absence of the
quantum-field-independent counterterms. The counterterms required to render
the theory finite were first classified in JO. They consist of all possible
diff-invariant dimension-four operators constructed out of the metric and
couplings and their derivatives:
\eqn{ W_{\text{c.t.}}=
-\int\sqrt{-\gamma}\,\mu^{-\epsilon}\lambda\cdot\mathscr{R},}
where dimensional regularization is used with $d=4-\epsilon$ and
\eqna{\lambda\cdot\mathscr{R}&=\lambda_aF+ \lambda_bG+
\lambda_cH^2+\mathscr{E}_i\partial_\mu g^i\partial^\mu H+\tfrac12
\mathscr{F}_{ij}\partial_\mu g^i\partial^\mu g^j H +\tfrac12
\mathscr{G}_{ij}\partial_\mu g^i\partial_\nu g^j G^{\mu\nu}\\ &\quad
+\tfrac12 \mathscr{A}_{ij}\nabla^2 g^i\nabla^2
g^j+\tfrac12\mathscr{B}_{ijk}\partial_\mu g^i\partial^\mu g^j\nabla^2 g^k
+\tfrac14 \mathscr{C}_{ijkl}\partial_\mu g^i \partial^\mu g^j \partial_\nu
g^k\partial^\nu g^l.}[localCTs]
Here $F$ is the Weyl tensor squared, $G$ is the Euler density, $H$ is
proportional to the Ricci scalar, and $G_{\mu\nu}$ is the Einstein tensor:
\begin{gather*}
  F=R^{\mu\nu\rho\sigma}R_{\mu\nu\rho\sigma}-\frac{4}{d-2}R^{\mu\nu}
  R_{\mu\nu}+\frac{2}{(d-2)(d-1)} R^2,\\
  G=\frac{2}{(d-3)(d-2)}(R^{\mu\nu\rho\sigma}R_{\mu\nu\rho\sigma}
  -4R^{\mu\nu}R_{\mu\nu}+R^2),\\
  H=\frac{1}{d-1} R,\qquad
  G_{\mu\nu}=\frac{2}{d-2}(R_{\mu\nu}-\tfrac12\gamma_{\mu\nu}R).
\end{gather*}
The quantities above are defined as in JO, with inessential $d$-dependent
factors for later convenience. Each of the counterterms in
$\lambda\cdot\mathscr{R}$ is an expansion in $1/\epsilon$ chosen to render
$\widetilde W$ finite---for this one must in addition introduce
wave-function and coupling constant counterterms, as usual. The
coefficients $\lambda=(\lambda_a,\lambda_b,\ldots,\mathscr{C}_{ijkl})$ are
in general functions of the couplings $g^i(x)$.

The anomalous variation of the generating functional is dictated by these
counterterms. While $W$ satisfies \LPRinvariance and \FULLinv, this is not
true of $W_{\text{c.t.}}$, as can be seen by explicit computation. The
anomaly is precisely the statement that the infinitesimal transformation
$\tau\to\tau+\sigma$ in \FULLinvb,
\eqn{\Delta_\sigma W_{\text{c.t.}}=
W_{\text{c.t.}}[(1-2\sigma)\gamma_{\mu\nu},
g^i-\sigma\mu \,dg^i/d\mu]-  W_{\text{c.t.}}[\gamma_{\mu\nu}, g^i],}
fails to vanish. The anomalous variation can be split into a term that
would occur even if $\sigma$ were spacetime-independent plus a term
proportional to the derivative of $\sigma$:
\eqn{\Delta W_{\text{anomaly}}=\Delta_\sigma W_{\text{c.t.}}=-\int
\sqrt{-\gamma}\,(\sigma\beta_\lambda\cdot\mathscr{R}
+\partial_\mu\sigma\,\mathscr{Z}^\mu).}[Wanomaly]
These terms again can be expanded using dimensional analysis and
diff-invariance:
\eqna{\beta_\lambda\cdot\mathscr{R} &=\beta_aF+\beta_bG+\beta_cH^2
+\chi^e_i\partial_\mu g^i\partial^\mu H+\tfrac12
\chi^f_{ij}\partial_\mu g^i\partial^\mu g^j H +\tfrac12
\chi^g_{ij}\partial_\mu g^i\partial_\nu g^j
G^{\mu\nu}\\
&\quad +\tfrac12 \chi^a_{ij}\nabla^2 g^i\nabla^2
g^j+\tfrac12\chi^b_{ijk}\partial_\mu g^i\partial^\mu g^j\nabla^2 g^k
+\tfrac14 \chi^c_{ijkl}\partial_\mu g^i \partial^\mu
g^j \partial_\nu g^k\partial^\nu g^l,}[betalambdadefined]
and\footnote{The second term involves the function of coupling constants
$d$, which is not to be confused with $d=4-\epsilon$. We follow Osborn in
this unfortunate choice of notation, hoping that with this warning no
confusion will arise in what follows.}
\eqna{\mathscr{Z}_\mu&=G_{\mu\nu}w_i\partial^\nu g^i +
\partial_\mu(Hd)+HY_i\partial_\mu g^i\\
&\quad +\partial_\mu(U_i\nabla^2 g^i+\tfrac12 V_{ij}\partial_\nu g^i
\partial^\nu g^j)+S_{ij}\partial_\mu g^i\nabla^2 g^j+\tfrac12
T_{ijk}\partial_\nu g^i\partial^\nu g^j\partial_\mu g^k,}[zetamudefined]
up to terms with vanishing divergence. Since $\widetilde W$ is finite and
the $\sigma$-variation of $W$ vanishes, it must be that the variation of
$W_\text{c.t.}$ is finite by itself.

Calculations of the coefficients in $\beta_\lambda\cdot\mathscr{R}$ and
$\mathscr{Z}_\mu$ can be done using standard techniques of dimensional
regularization with a mass-independent renormalization scheme, say MS.  For
now, let us concentrate on the relatively straightforward computation of
$\beta_\lambda\cdot\mathscr{R}$. Since for constant $\sigma$ the
transformation $\delta\gamma_{\mu\nu}=-2\sigma\gamma_{\mu\nu}$ just counts
dimensions, and the dimension of the volume element is $d$ while that of
the operators in $W_{\text{c.t.}}$ is four, we obtain
\eqn{(\epsilon-\hat{\beta}^i\hpartial_i)\lambda\cdot\mathscr{R}=
\beta_\lambda\cdot\mathscr{R}, }[RGEparameters]
where the beta function is
\eqn{\mu \frac{dg^i}{d\mu}=
\hat{\beta}^i=-\epsilon k^ig^i+\beta^i(g)\qquad\text{(no sum over }
i\text{)}.}
Here the derivative is taken holding the bare parameters fixed.  $k^i$ is
defined by requiring that the Lagrangian scales appropriately: for
$\phi'=\mu^{\delta\epsilon}\phi$ and $g^\prime{}^{i}=\mu^{k^i\epsilon}g^i$,
then $\mathscr{L}(\phi^\prime
,g^\prime)=\mu^{-\epsilon}\mathscr{L}(\phi,g)$. Note that
$\hat{\beta}^i\hpartial_i\equiv\hat{\beta}^i\hpartial/\hpartial g^i$
denotes substitution of $g^i$ by $\beta^i$ wherever $g^i$ may be found,
e.g., $\hat{\beta}^i\hpartial_i(\partial_\mu g^j)\equiv \partial_\mu
\beta^j=\partial_i\beta^j\partial_\mu g^i$, and of course respects the
standard rules of differentiation. Using \RGEparameters, it is
straightforward to show that, e.g.,
\eqn{\chi^a_{ij}=(\epsilon-\hat{\beta}^k\partial_k)\mathscr{A}_{ij}
-\mathscr{A}_{ik}\partial_j\hat{\beta}^k-\mathscr{A}_{jk}
\partial_i\hat{\beta}^k.}

The consistency conditions of JO can be understood as following from
requiring that \FULLinv applied to the complete renormalized generating
function $\widetilde W$ fails only up to the finite, anomaly terms.  The
left-hand side of \FULLinva does not involve any counterterms from
spacetime-dependent couplings, while the right-hand side does not involve
any from a curved background. Hence, the counterterms in one and the other
case must be related. Consider on the right-hand side of \FULLinva, for
example, the counterterm
\eqn{\tfrac12\mathscr{A}_{ij}\nabla^2 g^i \nabla^2 g^j=
\tfrac12\mathscr{A}_{ij}\hat\beta^i\hat\beta^j(\nabla^2 \tau)^2+\cdots,
}[Act]
where we have expanded to lowest order in $\tau(x)$. Comparing with the
counterterms on the left-hand side of \FULLinva, that arise solely from a
curved background, we have,
\eqn{\sqrt{-\gamma}\,(\lambda_a F+\lambda_b G
+ \lambda_c H^2)= 8\lambda_b\left[ (\nabla^2\tau)^2
-(\partial_\mu\partial_\nu\tau)^2+\cdots\right]
+4\lambda_c\left[(\nabla^2 \tau)^2+\cdots\right].}[GHtwovariation]
The $\lambda_b$ term is a total derivative so for localized $\tau(x)$ it
does not contribute (recall there is an implicit spacetime integration).
Matching the terms in~\Act and~\GHtwovariation we find that the
counterterms are related,
\eqn{4\lambda_c \sim \tfrac12\mathscr{A}_{ij}\hat\beta^i\hat\beta^j,}[cVSA]
where the symbol $\sim$ denotes equality up to finite terms, that is, the
difference is finite as $\epsilon\to0$. This precisely corresponds to
Eq.~(3.12) of JO. Applying $\mu \,d/d\mu$ on the bare couplings to derive
RGEs and the corresponding beta functions, one then derives from this JO's
consistency condition (3.21a),
\eqn{8\beta_c=\chi^a_{ij}\beta^i\beta^j-\beta^i\partial_i X,}[betacJO]
where $X$ arises from the finite difference between the left- and
right-hand sides of \cVSA, and $\beta_c$ and $\chi^a_{ij}$ are beta
functions for $\lambda_c$ and $\mathscr{A}_{ij}$, respectively. The
remaining consistency conditions in JO can be obtained in a similar
fashion. We only quote here one other consistency condition that plays an
important role in what follows. Using \FULLinva the lowest order terms in
$\tau(x)$ that are linear in the Einstein tensor give
\eqn{8\partial_i \lambda_b\sim \mathscr{G}_{ij}\hat \beta^j.}[dbVSG]
With the finite difference between the two sides of \dbVSG denoted by $w_i$
one obtains
\eqn{8\partial_i\beta_b= \chi^g_{ij}\beta^j-\beta^j\partial_j
w_i-\partial_i \beta^j w_j.}
This consistency condition is the origin of the proposal in JO for a
$c$-function,
\eqn{\tilde\beta_b\equiv \beta_b+\tfrac18\beta^iw_i,}
which satisfies
\eqn{\partial_i\tilde \beta_b= \tfrac18(\chi^g_{ij}+\partial_{[i}
w_{j]})\beta^j,}[JOWeylconditionA]
where $\partial_{[i}w_{j]}=\partial_iw_j-\partial_jw_i$. Then its RG flow
is monotonic provided the ``metric'' $\chi^g_{ij}$ is positive-definite,
for
\eqn{-\frac{d\tilde\beta_b}{dt}=\beta^i\partial_i\tilde \beta_b
= \tfrac18\chi^g_{ij}\beta^i\beta^j.}

To summarize, the extension \FULLinv of the invariance requirement of LPR
in \LPRinvariance, when applied to {\it the complete set of counterterms
required for finiteness when coupling constants have spacetime dependence},
leads to the consistency conditions of JO.

\subsec{The trace anomaly and the computation of
\texorpdfstring{$\nabla_\mu\mathscr{Z}^\mu$}{nabla Z}}[sec:Zmu]
As formulated, the renormalized generating functional $\widetilde W$ is a
finite function of the background metric and of renormalized
spacetime-dependent coupling constants. As such we can obtain finite
insertions of composite operators in Green
functions by functional differentiation,
\eqn{\vev{T_{\mu\nu}(x)}=\frac{2}{\sqrt{-\gamma}}\;
\frac{\delta\widetilde W}{\delta\gamma^{\mu\nu}(x)}
\qquad\text{and}\qquad
\vev{[\CO_i(x)]}=
\frac{1}{\sqrt{-\gamma}}\;
\frac{\delta\widetilde W}{\delta g^i(x)}~.}[Tmunudefined]
Note that $[\CO_i(x)]$ stands for the fully renormalized insertion of the
composite operator $\CO_i(x)$, which may differ from the operator monomial
in an expectation value. Following JO, we make this distinction explicit by
introducing the notation $[\ldots\xspace]$.

Using \Tmunudefined in \FULLinv and \Wanomaly one obtains
\eqn{T^\mu_{\phantom{\mu}\!\mu}=\hat\beta^i[\CO_i]-\mu^{-\epsilon}
\beta_\lambda\cdot\mathscr{R}
+\mu^{-\epsilon}\nabla_\mu\mathscr{Z}^\mu.}[traceanomaly]
This is the well-known trace anomaly, accounting for the effects of curved
background and spacetime-dependent coupling constants.  However, this
equation is not quite correct in the most generality: there are two terms
missing on the right-hand side. The first is an operator that vanishes by
the equations of motion times the anomalous dimension of the corresponding
quantum field. We have lost track of this term because the relation
\LPRinvariance is only correct up to terms that vanish by the equations of
motion. The second missing term is more subtle: we have missed counterterms
that may be needed to render some theories finite. When the kinetic terms
of the Lagrangian exhibit a continuous symmetry the current associated with
this symmetry is a dimension-three operator and a new type of counterterm
is required in the presence of spacetime-dependent couplings, that is, a
counterterm proportional to the product of the current and the derivative
of a coupling. This will be discussed extensively, and the anomaly equation
will be fixed, in Section \ref{Sec:DimThree}.

Let us turn to the computation of $\mathscr{Z}^\mu$ in \Wanomaly. It
follows, of course, straightforwardly from the definition \Wanomaly.
Slightly less trivial is the fact that the computation must give a finite
current $\mathscr{Z}^\mu$. That this must be so can be seen from the trace
anomaly in \traceanomaly, in which all other terms are already finite.
This means that there must be cancellations among infinite terms that
contribute to $\mathscr{Z}^\mu$. In fact, these cancellations are nothing
but the consistency conditions, e.g., \cVSA and \dbVSG. For example, the
terms in \Wanomaly involving the Einstein tensor (modulo terms that do not
vanish for spacetime-independent $\sigma$) are
\eqna{\int\sqrt{-\gamma}\,G^{\mu\nu}(-8\lambda_b\nabla_\mu\partial_\nu
\sigma-\mathscr{G}_{ij}\partial_\mu g^i \hat\beta^j\partial_\nu\sigma)
&=\int\sqrt{-\gamma}\,\partial_\nu\sigma\, G^{\mu\nu}\partial_\mu g^i
(8\partial_i\lambda_b-\mathscr{G}_{ij}\hat\beta^j)\\
&=\int\sqrt{-\gamma}\,\partial_\nu\sigma\, G^{\mu\nu}\partial_\mu g^i\,
w_i.}
Here, in going from the first to the second line we used the finiteness
condition \dbVSG and the definition that the finite difference is $w_i$.
Thus we have reproduced the first term in $\mathscr{Z}_\mu$ of
\zetamudefined. The remaining terms in \zetamudefined can be similarly
found.

\subsec{Wess--Zumino action}[sec:WZ]
The Wess--Zumino action, $W_\text{WZ}$, is some function of $\tau(x)$ that
will give $-\Delta  W_\text{anomaly}$ upon a Weyl transformation,
$\tau(x)\to\tau(x)+\sigma(x)$.  Focusing on the $\beta_b$-term in $\Delta
W_{\text{anomaly}}$,
\eqn{-\int\sqrt{-\gamma}\,\sigma\beta_b G,}[sigmabetaG]
KS write the corresponding WZ term as\foot{The sign in the term cubic in
$\tau$ is opposite to that of KS because we use the opposite sign
convention for the conformal compensator, which gives $\Delta_\sigma$ as in
JO.}
\eqn{\int\sqrt{-\gamma}\left\{\tau \beta_b G-4\beta_b
    \left[G^{\mu\nu}\tau_{,\mu}\tau_{,\nu}
      +\tau_{,\mu}\tau^{,\mu}\nabla^2\tau
      +\tfrac12(\tau_{,\mu}\tau^{,\mu})^2\right]\right\},}[KSWZ]
where we have introduced the shorthand $\tau_{,\mu}=\partial_\mu\tau$.
However, this computation is incomplete. The problem with this is that we
have ignored the effect of the new counterterms arising from spacetime
dependence of the couplings. Since we will not need a Wess--Zumino action
for our generalization of the KS argument to theories with cycles, we will
not aim at being complete and only point out one interesting consequence
here.  Consider, for example, the term in $\mathscr{Z}^\mu$
\eqn{-\int\sqrt{-\gamma}\,\partial_\mu\sigma\,w_i
  G^{\mu\nu}\partial_\nu g^i.
}[sigmaWGdg]
Now with $\partial_\mu g^i=-\beta^i\tau_{,\mu}$, one has the following
generalization of the Wess--Zumino dilaton action:
\eqn{\int\sqrt{-\gamma}\left\{\beta_b\tau G
-4(\beta_b+\tfrac18w_i\beta^i) \left[G^{\mu\nu}\tau_{,\mu}\tau_{,\nu}
+\tau_{,\mu}\tau^{,\mu}\nabla^2\tau
+\tfrac12(\tau_{,\mu}\tau^{,\mu})^2\right]\right\}.}[WZimproved]
The Weyl variation of \WZimproved gives the sum of~\sigmabetaG
and~\sigmaWGdg (if $\partial_\mu g^i=-\beta^i\tau_{,\mu}$ there).  The
correction that takes $\beta_b$ into
$\tilde\beta_b=\beta_b+\tfrac18w_i\beta^i$ is generally of lower order than
the running of $\beta_b$. That is, $\tfrac18w_i\beta^i$ is of lower order
than $\beta_b-\beta_{b0}$, where $\beta_{b0}$ stands for the free field
theory value of $\beta_b$.

Let us be more explicit. Consider, for example, the perturbative result of
JO for a pure YM theory with gauge coupling $g$,
\eqn{\beta_b=\beta_{b0}+ \frac{n_V\beta_1}{8(16\pi^2)^3}g^4
+\CO(g^6),}[betabtwoloops]
from which $\beta_b$ is seen to increase in the flow out of the trivial UV
fixed point. JO also give $gw=2n_V/16\pi^2+\cdots$, and therefore
\eqn{\tilde\beta_b= \beta_{b0} -
\frac{n_V\beta_0}{4(16\pi^2)^2}g^2+\CO(g^4),}
which shows that the leading-order running of $\tilde \beta_b$ is modified
by the $\tfrac18w\beta^g$ term. Note that $\tilde\beta_b$ decreases in the
flow out of the trivial UV fixed point, as opposed to $\beta_b$ which, as
seen from \betabtwoloops, increases. Therefore, a strong $c$-theorem in
four dimensions should involve $\tilde\beta_b$, not $\beta_b$. Of course
$\tilde\beta_b$ and $\beta_b$ agree at fixed points.

There is another subtle point we would like to address. The coefficient
appearing in the two-point function of the trace of the stress-energy
tensor appears to play the role of the ``metric'' $\chi^g_{ij}$ in the
consistency condition \JOWeylconditionA. In Appendix~\ref{app:chigvschia}
we point out, following JO, that this is actually related to
$-2\chi^a_{ij}$, see \betacJO. Explicit computations show that
$-2\chi^a_{ij}$ agrees with $\chi^g_{ij}$ to second order in perturbation
theory in any four-dimensional theory. As we show in
Appendix~\ref{app:chigvschia} this agreement fails, for example, at third
order in a YM theory with a single gauge coupling.

\newsec{Flavor symmetries, dimension-three operators and the corrected
trace anomaly}[Sec:DimThree]
As we have mentioned above, in deriving the trace anomaly we have missed
counterterms that may be needed to render some theories finite. When the
kinetic terms of the Lagrangian exhibit a continuous symmetry the current
associated with this symmetry is a dimension-three operator and a new type
of counterterm is required in the presence of spacetime-dependent
couplings, that is, a counterterm proportional to the product of the
current and the derivative of a coupling.

Consider a theory with $n_S$ real scalar fields interacting through the
usual quartic interaction. The kinetic part of the bare Lagrangian,
\eqn{\mathscr{L}_{0K}=\tfrac12\partial^\mu\phi_{0a}\partial_\mu\phi_{0a},}
exhibits a continuous symmetry under transformations of the fields
$\delta\phi_{0a}=-\omega_{ab}\phi_{0b}$, where $\omega$ is in the algebra
of the flavor group $G_F=SO(n_S)$. In the process of renormalization we
introduce a renormalization matrix $Z$ and write
\eqn{\mathscr{L}_{0K}=\tfrac12\partial^\mu\phi^TZ\partial_\mu\phi,}
where renormalized fields, $\phi$, are related to bare fields by
$\phi_0=Z^{1/2}\phi$.\foot{Note that in this step we have the freedom to
introduce an orthogonal matrix $O$ and define $\phi_0=\tilde Z^{1/2}\phi$,
where $\tilde Z^{1/2}=OZ^{1/2}$. This does not affect $Z=Z^{T/2}Z^{1/2}$.
Nevertheless, such a freedom leads to an ambiguity in the definition of
beta functions and anomalous dimensions as we explain in Appendix
\ref{app:Bfunction}.}  In the presence of spacetime-dependent coupling
constants new divergences arise and thus new counterterms are needed. For
example, one must introduce a new counterterm of the form
\eqn{\mathscr{L}_{\text{c.t.}}=(\partial^\mu g^i)
(N_i)_{ab}\phi_{0b}\partial_\mu\phi_{0a},}[prenewct]
with $(N_i)_{ab}=-(N_i)_{ba}$, that is, in the algebra of $G_F$. Note that
this new counterterm is not accounted for in $W_{\text{c.t.}}$ which by
construction is independent of quantum fields.  Note also that additional
counterterms, symmetric under $a\leftrightarrow b$, must also be
introduced.  One may integrate by parts to write these as terms with no
derivatives acting on the quantum fields. While necessary, they do not play
a central role in what follows.

To be more explicit, we consider a theory of real scalars and write for the
bare Lagrangian
\eqn{\mathscr{L}_0=\tfrac12 \gamma^{\mu\nu}D_{0\mu}\phi_{0a}D_{0\nu}\phi_{0a}
+\tfrac18(d-2)\phi_{0a}\phi_{0a} H -
\tfrac{1}{4!}g^0_{abcd}\phi_{0a}\phi_{0b}\phi_{0c}\phi_{0d}.
}[barescalarlag]
This is written in term of bare fields $\phi_0$. The second term is
introduced to ensure conformal invariance of the classical action. In the
potential term, the bare couplings $g^0_{abcd}$ are completely symmetric
under exchange of the indices $a, b, c$ and $d$. The covariant derivative,
\eqn{D_{0\mu}\phi_0=(\partial_\mu +A_{0\mu})\phi_0,}
is introduced with an eye towards including the counterterm \prenewct,
since
\eqn{A_{0\mu}=A_{\mu}+N_I(D_\mu g)_I,\qquad D_\mu=\partial_\mu+A_\mu.
}[bareAdefined]
Here, following JO, we use the compact notation $I=(abcd)$ and we have left
implicit the Lie-algebra indices (so that $N_I^T=-N_I$ and
$A_\mu^T=-A_\mu$). Note that $N_I$ is a function of the renormalized
couplings that has an $\epsilon$-expansion starting at order $1/\epsilon$.
If the theory contains gauge fields and some of the scalars are charged
under the gauge group $G_g\subseteq G_F$, it is straightforward to include
an additional quantum gauge field in addition to the background field
$A_\mu$.

The Lagrangian \barescalarlag is explicitly locally $G_F$-symmetric if we
agree to transform the couplings and the gauge fields:
\eqna{\delta g_{abcd}^0&=-\omega_{ae}g^0_{ebcd}+\text{permutations} \qquad
(\delta g^0_I =-(\omega g^0)_I~\text{for short}),\\ \delta A_\mu &=
D_\mu\omega.}
The first of these is already used in defining the covariant derivative
$(D_\mu g)_I$ in \bareAdefined. It is very important to note at this point
that if this explicit local invariance is non-anomalous it can (and will)
be used to constrain the counterterms and the generating functional
$\widetilde W$,
\eqn{\widetilde W[\gamma^{\mu\nu}(x),(\Omega g)_I(x),\Omega D_\mu\Omega^{-1}]=
\widetilde W[\gamma^{\mu\nu}(x),g_I(x),A_\mu],}[WGFinv]
where $\Omega(x)=\exp(\omega(x))\in G_F$.  Of course, in theories without
spinors the symmetry is trivially non-anomalous. Furthermore, derivatives of
the generating functional with respect to the background field now give
insertions of the scalar current.

It is not our intention to repeat the calculations of JO in their entirety
here. We will instead describe the main ingredients and results. We have
already described the two main new ingredients, namely, the need for new
counterterms and the introduction of a background field to ensure
invariance under $G_F$ in \WGFinv. As before, additional
quantum-field-independent counterterms are required. These are as in
\localCTs but with the replacement $\partial_\mu\to D_\mu$ to ensure $G_F$
invariance.  Additional counterterms involving the field strength
$F_{\mu\nu}=[D_\mu,D_\nu]$ are also required,
\eqn{\tilde\lambda\cdot \mathscr{R}=\lambda\cdot \mathscr{R}
+\tfrac14\text{Tr}(\mathscr{K}F^{\mu\nu}F_{\mu\nu})+
\tfrac12\text{Tr}(\mathscr{P}_{IJ}F^{\mu\nu})(D_\mu g)_I(D_\nu g)_J.
}[newCTs]
Moreover, as advertised, new field-dependent counterterms must be included,
\eqn{\mathscr{Q}=\eta_{ab}\phi_a\phi_b
H+(\delta_I)_{ab}\phi_a\phi_b(D^2g)_I+\tfrac12
(\epsilon_{IJ})_{ab}\phi_a\phi_b (D^\mu g)_I(D_\mu g)_J.}[QCTs]

Proceeding much as before, JO find\foot{Note that in Jack and Osborn the
first term contains $\hat\beta^V=\mu\,dV/d\mu$, where $V$ is the
renormalized potential, and with the derivative taken by holding the bare
fields, $\phi_0$, and the bare potential, $V_0$, constant (independent of
RG time). With a potential of the form $V=g_I\CO_I$ and following Jack and
Osborn's definitions we then obtain $[\hat\beta^V]=\hat\beta_I[\CO_I]$ as
expected.} \cite[Eq.~(6.15)]{Jack:1990eb}
\eqn{T^\mu_{\phantom{\mu}\!\mu}
=\hat \beta_I[\CO_I] + [\beta^{\mathscr{Q}}] + [(D^\mu\phi)^T\beta_\mu^A\phi]
-\mu^{-\epsilon}\beta_{\tilde\lambda}\cdot\mathscr{R}
+\nabla_\mu (J^\mu +J_\Theta^\mu + \widetilde{\mathscr{Z}}^\mu)
-((1+\hat\gamma)\phi)\cdot\frac{\delta}{\delta\phi}\tilde
S_0,}[TmumuGeneralJmu]
which, using the underlying gauge invariance, they rewrite
as~\cite[Eq.~(6.23)]{Jack:1990eb}
\eqn{T^\mu_{\phantom{\mu}\!\mu}
=\widehat B_I[\CO_I]+[\beta^{\mathscr{Q}}] + [(D_\mu\phi)^TB_\mu^A\phi]
-\mu^{-\epsilon}\beta_{\tilde\lambda}\cdot\mathscr{R}
+\nabla_\mu (J_\Theta^\mu + \widetilde{\mathscr{Z}}^\mu)
-((1+\hat\gamma+S)\phi)\cdot\frac{\delta}{\delta\phi}\tilde
S_0.}[TmumuGeneral]
Many comments are in order. The last term, involving the derivative of the
full action integral $\tilde S_0$, vanishes by the equations of motion. We
have included it here for completeness. We have already commented that a
similar term is missing from \traceanomaly. The operator $[\CO_I]$
corresponds to the interaction term in the Lagrangian,
$\CO_I=\frac1{4!}\phi_a\phi_b\phi_c\phi_d$, but differs from it,
$[\CO_I]=\CO_I-\nabla_\mu J_I^\mu$ where
$J_I^\mu=(D_0^\mu\phi_0)^TN_I\phi_0$. Its coefficient in~\TmumuGeneral
is given by
\eqn{\widehat B_I\equiv \hat \beta_I- (Sg)_I,}[Bconf]
where $\hat\beta_I=\mu \,dg_I/d\mu=-\epsilon g_I+\beta_I$.  The current
$J^\mu$ in \TmumuGeneralJmu is defined as
\eqn{J^\mu=(D_0^\mu\phi_0)^TN_I\hat\beta_I\phi_0}[Jdefd]
and it is finite as required from consistency of \TmumuGeneralJmu. Note
that the combination $[\CO_I]+\nabla_\mu J^\mu_I$ appearing in
\eqref{TmumuGeneralJmu} is just $\CO_I$. However, $\CO_I$ is not by itself
a finite operator. While $\hat\beta_I\CO_I$ is finite, since it is the sum
of two finite operators, replacing $\hat\beta_I$ by its  $\epsilon\to0$
limit, $\beta_I\CO_I$, is not by itself finite. Finiteness of $J^\mu$
implies that it can be brought to the form
\eqn{J^\mu=[(D^\mu\phi)^TS\phi].}
The Lie-algebra element $S$ is then defined by $\widehat B_I N_I=S$. Since
$S$ is finite it is required that the infinite pieces of $\widehat B_I N_I$
cancel automatically, i.e.,
\eqn{\widehat B_I N_I = S\,\Rightarrow\, S=-N_I^1g_I,}[Sdef]
where $N_I=\sum_{n=1}^\infty N^n_I/\epsilon^n$, so that $N_I^1$ is the
residue of the simple $\epsilon$-pole in $N_I$. Cancellation of the
infinite pieces requires that $B_IN_I^n-g_IN_I^{n+1}=0$ for $n\ge1$. The
beta functions for the field-dependent quadratic counterterms are
\eqn{\beta^{\mathscr{Q}}\equiv
\beta^\eta_{ab}\phi_a\phi_b H
+(\beta^\delta_I)_{ab}\phi_a\phi_b(D^2g)_I
+\tfrac12(\beta^\epsilon_{IJ})_{ab}\phi_a\phi_b (D_\mu g)_I(D_\mu g)_J\,.}
The term $\beta_{\tilde\lambda}\cdot \mathscr{R}$ is the obvious
generalization of \betalambdadefined while the current
${\widetilde{\mathscr{Z}}}^\mu$ is defined as in \zetamudefined but
rendered $G_F$-invariant by replacing derivatives by covariant derivatives.
In addition, $\widetilde{\mathscr{Z}}^\mu$ has contributions from the new
counterterms in \newCTs, and there are additional contributions to the
terms with the $\mathscr{A}$ and $\mathscr{B}$ of \localCTs. The third term
in \TmumuGeneral involves
\eqn{B^A_\mu \equiv \beta^A_\mu+D_\mu S \equiv \rho_I (D_\mu g)_I + D_\mu
S\equiv P_I (D_\mu g)_I,\qquad \rho_I=g_J\partial_J N_I^1+N_I^1,}
where $\beta^A_\mu\equiv \mu \,dA_\mu/d\mu$ is the beta function for the
background gauge field $A_\mu$.  Finally, the current $J^\mu_\Theta$ arises
from the counterterms in \QCTs and has a complicated expression in terms of
the simple $\epsilon$-poles in $\delta_I$ and $\epsilon_{IJ}$ (see JO for
details \cite[Eqs.~(6.21--22)]{Jack:1990eb}).

At this point we can take the limit of flat spacetime,
spacetime-independent couplings and no background gauge field in
\TmumuGeneral. This gives
\eqn{T^\mu_{\phantom{\mu}\!\mu}=\widehat B_I[\CO_I]-
((1+\hat\gamma+S)\phi)\cdot\frac{\delta}{\delta\phi}S_0.}[traceT]
Since $[\CO_I]$ is finite we can now safely conclude that a theory is
conformal if and only if $B_I=0$. This does not require that $\beta_I=0$.

In the general case considered here the JO consistency conditions are
modified relative to what has been presented in Section \ref{Sec:WeylCC}.
On the one hand the conditions have to be covariant under transformations
by the symmetry group $G_F$. On the other, there are additional terms that
arise from the additional counterterms required to render the theory
finite. Osborn gives the form of these most general consistency conditions
\cite{Osborn:1991gm}. Two conditions play a role in our discussion:
\eqna{8\partial_I\beta_b &=\chi^g_{IJ} B_J-B_J\partial_J w_I-
(\partial_I B_J) w_J - (P_I g)_J w_J\\
&=\chi^g_{IJ} B_J-\beta_J\partial_J w_I-
(\partial_I \beta_J) w_J - (\rho_I g)_J w_J,}[betabGenCC]
and
\eqn{B_IP_I=0.}[BIPICC]
In addition, covariance under $G_F$ gives, e.g.,
\eqn{(\omega g)_I \partial_I\beta_b =0\qquad\text{and}\qquad
(\omega g)_I \partial_I S = [\omega,S].}[GFCC]
Of course, the first of these applies to any $G_F$-invariant while the
second to any antisymmetric tensor (for example, any Lie-algebra valued
function).  Using the first of \GFCC in \betabGenCC gives a nontrivial
relation among several beta functions:
\eqn{(\omega g)_I\left[\chi^g_{IJ} B_J-B_J\partial_J w_I-
(\partial_I B_J) w_J - (P_I g)_J w_J\right]=0,}[cancelA]
or, equivalently,
\eqn{
(\omega g)_I\left[\chi^g_{IJ} B_J-\beta_J\partial_J w_I-
(\partial_I \beta_J) w_J - (\rho_I g)_J w_J\right]=0.
}[cancelB]

These conditions can be used to understand aspects of the flow of
$\beta_b$. Consider the flow defined by some arbitrary function $f_I(g)$,
\eqn{\frac{d\bar g_I}{d\eta}= -f_I(\bar g(\eta)).}
If one takes $f_I=\beta_I$ then the flow can be identified with the RG
flow, with $\eta=t=-\ln(\mu/\mu_0)$. From \betabGenCC we have
\eqn{-8\frac{d\widetilde{B}_b}{d\eta}=\chi^g_{IJ}f_I B_J
+f_I B_J\partial_{[I} w_{J]} - (P_I g)_J f_I w_J,}[etaflowwithB]
where
\eqn{\widetilde{B}_b=\beta_b+\tfrac18 B_Iw_I,}[Btildeb]
and
\eqn{-8\frac{d\tilde \beta_b}{d\eta}=\chi^g_{IJ}f_I B_J
+f_I\beta_J\partial_{[I} w_{J]} - (\rho_I g)_J f_I w_J.}[etaflowwithbeta]
Three special cases are of most interest. Consider first $f_I(g)=-(\omega
g)_I$.  From the second equation in \GFCC we see that on this flow $\omega$
is constant.  This is a recursive flow (cycle or ergodic). It follows from
the $G_F$-invariance of $\widetilde{B}_b$ and $\tilde\beta_b$ that these
remain constant on the flow. This is a consequence of the detailed
cancellations that must be satisfied by the beta functions in \cancelA and
\cancelB. This general result can be applied to  limit cycles,
$\beta_I=(Qg)_I$,  for which $\omega=Q$.  We thus see that counterterms
that ensure $G_F$-covariance guarantee constancy of $\beta_b$ (and
$\tilde\beta_b$) on recursive flows.

The second and third special cases correspond to $f_I=B_I$ and
$f_I=\beta_I$. While the $B_I$ and $\beta_I$ flows are generally
distinguishable, one may use \GFCC to show the two flows are identical for
$G_F$-invariants.\foot{This was pointed out to us by J.\ Polchinski.} Using
\etaflowwithB with $f_I=B_I$ and the consistency condition \BIPICC we see
that
\eqn{-8\frac{d\widetilde{B}_b}{dt}=\chi^g_{IJ}B_IB_J.}[Bflowofbeta]
This shows that $\widetilde B_b$ decreases monotonically along both flows
and is a good candidate for the $c$ function of the $c$-theorem. Indeed
this shows a strong version of the $c$-theorem in perturbation theory. To
two loops $\chi^g_{IJ}=-2\chi^a_{IJ}>0$, where unitarity is required for
the inequality, so the right-hand side of \Bflowofbeta is positive-definite
along a perturbative flow.

The relation between the $B_I$ and $\beta_I$ flows can be made more
explicit, hopefully clarifying their relation. Consider the flows
\eqn{-\frac{dg_I}{dt}=\beta_I(g(t))\qquad\text{and}\qquad -\frac{d\bar g_I}{d\eta}=B_I(\bar g(\eta)).}
The solution to the $\eta$-flow is given in terms of the one for the
RG flow by
\eqn{\bar g(\eta) =
  F(\eta)g(\eta)\qquad\text{where}\qquad F(\eta)=T\left(\exp\left[-\int_{-\infty}^\eta
      d\eta' S(\eta')\right]\right).}
Here $T$ is the $\eta$-ordered product and $F\in G_F$. As such,
$\beta_I(\bar g)=\beta_I(Fg)=(F\beta)_I(g)$ and similarly for $B_I$ and
indeed for any tensor function of the couplings. Of special interest are
$G_F$-invariants, like $\widetilde{B}_b$, for which $\widetilde{B}_b(\bar
g) = \widetilde{B}_b(Fg)=\widetilde{B}_b( g) $.  So   we see again that the
monotonic $\eta$-flow of $\widetilde{B}_b$ gives a monotonic RG flow of
$\widetilde{B}_b$.

The quantity $\tilde \beta_b$ does not appear to be a good candidate
for the $c$ function of the $c$-theorem.  Using  \etaflowwithbeta to study its flow, so
the term $f_I\beta_J\partial_{[I} w_{J]}$ automatically vanishes, we obtain
\eqn{-8\frac{d\tilde \beta_b}{dt}=\chi^g_{IJ}\beta_I B_J
 - (\rho_I g)_J \beta_I w_J.}[wouldbeCthm]
Were we to ignore the last term on the right-hand side we would be able to
establish a perturbative $c$-theorem for $\tilde\beta_b$. Indeed, to two
loops $B_I=\beta_I$ and $\chi^g_{IJ}=-2\chi^a_{IJ}>0$ so the right-hand
side of \wouldbeCthm would be positive-definite along a perturbative flow.
However, the last term is parametrically of the same order as the first on
the right-hand side of \wouldbeCthm so this does not give a perturbative
$c$-theorem for $\tilde\beta_b$.

\newsec{Scale implies conformal invariance}
\subsec{\texorpdfstring{$S$}{S} is \texorpdfstring{$Q$}{Q} (on
cycles)}[QvsS]
In this subsection we elucidate the relation between $Q$ and $S$. Our
treatment is focused on theories in $d=4$. We remind the reader that $Q$ is
defined as the solution to the equations $\beta^g=0$ and $\beta_I=(Qg)_I$,
defining an RG cycle on which $Q$ remains a constant while $S$ is defined
as a function of couplings that makes explicit the finiteness of the
current $J^\mu$ in \eqref{Jdefd}. There is no a priori reason they should
be related.

What is known about $S$? JO have shown, by direct calculation, and we have
verified, that in a scalar field theory $S$ vanishes up to third order in
the loop expansion. The result holds even if gauge fields are included and
the scalars are charged under the gauge group. For theories with scalars
and fermions, JO have shown, and we have verified, that $S$ remains zero to
two loops. However, this is consistent with a possible equality of $S$ and
$Q$ on cycles. Indeed, we have obtained previously that $Q$ is of third
order in the loop expansion in Yang--Mills theories with scalars and
fermions, while in purely-scalar field theories a non-vanishing $Q$, if it
exists, must be at least of fifth order in the loop expansion.

As might be expected from the discussion above, we will show that (up to
conserved current)
\begin{enumerate}
\item $S$ is $Q$ on cycles,
\item $S$ vanishes at fixed points.
\end{enumerate}
In light of these results the computation of $Q$ can be tremendously
simplified given an explicit expression for $S$. Presently, the procedure
to determine $Q$ involves determining first the beta functions for the
coupling constants to second order in the loop expansion for scalar
self-couplings, to third order in the loop expansion for Yukawa couplings
and to fourth order in the loop expansion for Yang--Mills couplings, and
then solving the system of nonlinear coupled equations $\beta^g=0$ and
$\beta_I=(Qg)_I$ (we implicitly use here that $g_I$ can also stand for
Yukawa couplings). Since $S$ must have a perturbative expansion that starts
at third order in the loop expansion, to determine $Q$ from $S$ it suffices
to evaluate it with coupling constants on the cycle computed to lowest
order in the loop expansion. So $Q$ is obtained from $S$ by determining the
zeroes of the one-loop beta functions (two-loop for gauge couplings): if
$S=0$ on the zero of the beta functions, the zero is a fixed point of the
RGE, but if $S\ne0$ on the zero, then the zero is a point on a cycle and
$Q=S$ there.

To this end an explicit, three-loop expression for $S$ is required.  But as
pointed out above, there has been no computation of $S$ to the order where
one would expect it to be non-vanishing if $S$ were to equal $Q$ on cycles.
We have endeavored to compute $S$ to third order in the loop expansion for
a general theory containing $n_S$ real scalars and $n_f$ Weyl spinors,
possibly charged under a gauge group.  The potential in the Lagrangian is
\eqn{V= \tfrac{1}{4!}\lambda_{abcd}\phi_a\phi_b\phi_c\phi_d
+(\tfrac12 y_{a|ij}\phi_a\psi_i\psi_j + \text{h.c.}).}
The details of the computation are spelled out in Appendix~\ref{app:S}.
The surprisingly simple result is
\eqn{(16\pi^2)^3S_{ab}=\tfrac{5}{8}\tr(y^{\phantom{*}}_ay^*_c
y^{\phantom{*}}_dy^*_e)\lambda_{bcde}+\tfrac{3}{8}
\tr(y^{\phantom{*}}_ay^*_cy^{\phantom{*}}_dy^*_dy^{\phantom{*}}_by^*_c)
-\{a\leftrightarrow b\}+{\rm h.c.}.}
We have evaluated this expression on the fixed points and cycles of the
theories we explored in \cite{Fortin:2011ks, Fortin:2012ic, Fortin:2012cq}
and found that in each case, even in examples in $d=4-\epsilon$, $S$
vanishes at all fixed points and equals our previous determination of $Q$
on all cycles.

Now for the (perturbative) proof of the propositions above. First we show
that $S=Q$ on cycles. Consider the $\eta$-flow with $f_I=B_I$, with
boundary condition that at $\eta=0$ the point $\bar g_I(0)$ is on the
cycle. Then $B_I(0)=\beta_I(\bar g(0))-(Sg(0))_I=([Q-S]g(0))_I$, with $Q-S$
in the Lie algebra of $G_F$ and the left-hand side of \Bflowofbeta
vanishes. Since $\chi^g_{IJ}$ is positive-definite to second order in the
loop expansion, \Bflowofbeta gives $B_I(0)=0$. This implies $S=Q+\Delta Q$
on cycles, where $(\Delta Q\,g)_I=0$. But if $\Delta Q\ne 0$ this
corresponds to a conserved current, $\nabla_\mu[(D^\mu\phi)^T\Delta
Q\;\phi]=0$, and we are free to redefine the scale current by a conserved
current by $Q\to Q+\Delta Q$.  Hence, $S=Q$ on cycles.\foot{In unitary
theories with $\mathcal N=1$ supersymmetry we recently showed, without
relying on perturbation theory, that $S=0$ \cite{Fortin:2012hc}. It thus
follows that RG limit cycles do not arise in such theories.}

For theories with two scalars there is an alternative, perhaps simpler
proof that $S$ equals $Q$ when evaluated on a cycle. Consider \Bflowofbeta
evaluated on a point on the cycle. It is easy to show that $S$ is a
constant on the cycle: $-dS/dt=\beta_I\partial_I S=(Qg)_I\partial_I
S=[Q,S]=0$, where the last two steps follow from \GFCC and the fact that,
for two flavors, the flavor group, $SO(2)$, is Abelian. Now, as before, we
consider the $\eta$-flow defined by the $B_I$ function starting from a
point on the RG-cycle (we make the distinction of the actual RG-cycle and a
$\eta$-cycle explicit, to avoid confusion). The flow is defined by $-d\bar
g_I /d\eta = B_I=\beta_I-(Sg)_I=([Q-S]g)_I$, where the last step follows
from assuming the initial point is on the RG-cycle and then noting that the
solution corresponds to a trajectory that traverses the same cycle but at a
different angular speed (the angular speeds are $Q_{12}$ and
$Q_{12}-S_{12}$ for the RG- and $\eta$-cycles, respectively). Therefore the
$\eta$-cycle is generated by a trajectory in $G_F$ and it follows that,
just as for an RG-cycle, any $G_F$-invariant remains constant on the
$\eta$-cycle. But the consistency condition \Bflowofbeta then implies that
$B_I=([Q-S]g)_I=0$ on the cycle. Since $Q$ and $S$ are each characterized
by a single number the only solution is $S_{12}=Q_{12}$ (on the cycle).

It is easy to show that $(Sg)_I=0$ at a fixed point, and this is consistent
with the notion that a fixed point corresponds to the case $(Qg)_I=0$. To
see this, notice that at a fixed point $B_I=-(Sg)_I$ so at that point the
flow corresponds to a first-order $G_F$-transformation. That is, the first
derivative with respect to $\eta$ of $G_F$-invariants vanishes at the fixed
point. Hence, \Bflowofbeta gives that $\chi^g_{IJ}(Sg)_I(Sg)_J=0$ and hence
$(Sg)_I=0$ at the fixed point. The solution is that either $S=0$ at the
fixed point, or there is an emergent symmetry at the fixed point, and
$J^\mu$ is the corresponding conserved current. This completes the proof of
the two propositions above.

\subsec{Cyclic CFTs}[cyclicCFTs]
\subsubsec{A perturbative proof that scale imples conformal invariance}
The condition for a theory in $d>2$ to be scale-invariant is that the trace
of its stress-energy tensor be a total derivative \cite{Polchinski:1987dy},
\eqn{T^\mu_{\phantom{\mu}\!\mu}=\partial_\mu V^\mu,}
where $V^\mu\neq j^\mu+\partial_\nu L^{\mu\nu}$ with $\partial_\mu j^\mu=0$
and, without loss of generality, $L^{\mu\nu}=L^{\nu\mu}$. A candidate for
$V^\mu$ is $V^\mu=\partial^\mu\phi^TP\phi$. If the theory includes spinors
an additional current can be added to $V^\mu$ but the argument below is
easily generalized by trivial extensions, e.g., by interpreting the index
$I$ as including all couplings. Using the equations of motion, or
alternatively a $G_F$-transformation, this can be cast as an algebraic
condition,
\eqn{B_I=(Pg)_I.}[condscale]
It is easy to see now that in $d=4$ the $B_I$-flow of $\widetilde B_b$
requires $(Pg)_I=0$. Indeed,  using \condscale in \Bflowofbeta the
left-hand side vanishes on account of $B_I$ being of the form $(\omega
g)_I$, and then perturbative positivity of $\chi^g_{IJ}$ implies $B_I=0$.
While $P$ may not vanish, the current $V^\mu$ can at most be a symmetry of
the theory, $V^\mu=j^\mu$. This concludes the proof that scale implies
conformal invariance in perturbation theory.

\subsubsec{Some properties of cyclic CFTs}
Our result that scale implies conformal invariance implies that the
non-trivial cycle found in \cite{Fortin:2012cq} actually corresponds to a
CFT. We dub such CFTs cyclic CFTs. It is quite surprising that CFTs can be
found at points where the beta functions do not vanish.  It is unclear
what, if anything, distinguishes these theories from fixed-point CFTs.
Presumably the special current $J^\mu$ plays a crucial role. We hope to
address these questions in the future, but at present have no progress to
report.

Since the stress-energy tensor is not renormalized, and since the
divergence of the special current $J^\mu$ appears in the trace-anomaly
equation, one may suspect its anomalous dimension vanishes. If so this
would correspond to a non-conserved vector operator of dimension exactly
three (no anomalous dimension), which is impossible in a unitary CFT.
However, the operator actually mixes under renormalization. A simple
computation gives
\eqna{\mu\frac{d}{d\mu}[\CO_I]&=
  -\partial_I\hat\beta_J[\CO_J]+\partial_\mu[\partial^\mu\phi^T\rho_I\phi],\\
\mu\frac{d}{d\mu}[\partial^\mu\phi^T\omega\phi]&=-[\partial^\mu\phi^T\rho_I(\omega
g)_I\phi],}
which allows one to readily verify that (i) the combination
$\hat\beta_I[\CO_I]+\partial_\mu J^\mu$ is RG-invariant, (ii) a
symmetry current is RG-invariant, and (iii) $J^\mu$ is not
RG-invariant,
$\mu\frac{d}{d\mu}J^\mu=-[\partial^\mu\phi^T\hat\beta_I\rho_I\phi]$.

Even if the beta function is non-vanishing, properties that follow directly
from the conformal symmetry apply to these cyclic CFTs. Consider for example
the well-known fact that two point correlators of primary operators can be
diagonalized and
\eqn{\vev{\CO(x)^\dagger\CO(0)}=(x^2)^{-\Delta_{\CO}}.}
Now contrast this with the two point function of the elementary real
scalars $\phi_a$ in a cyclic CFT. Scale and Poincar\'{e} invariance alone
give\cite{Fortin:2011bm}
\eqn{\vev{\phi(x)\phi^T(0)}=(x^2)^{-\frac12\Delta}G(x^2)^{-\frac12\Delta^T},}[twptfnctn]
where $G$ is a fixed real, positive, symmetric matrix and $\Delta = 1
+\gamma +Q$, with $\gamma^T=\gamma$ the anomalous dimension matrix of the
elementary fields $\phi$ and $Q^T=-Q$ defining the cycle through
$\beta_I=(Qg)_I$. Now one can redefine the field by $\phi\to M^{-1}\phi$
with $M$ chosen so that $MGM^T=1$, which is always possible with real $M$
for a real, positive, symmetric matrix. This effectively redefines
$\Delta\to M\Delta M^{-1}$. The condition for invariance under special
conformal transformations then gives\footnote{Alternatively, special
conformal transformations on \twptfnctn require that $\Delta G=G\Delta^T$.}
$\Delta^T=\Delta$. A further field redefinition by an appropriate
orthogonal transformation $R$ finally brings $\Delta$ into diagonal form,
$\Delta\to R\Delta R^T$. The entries of this diagonal form of $\Delta$
correspond to the roots of the characteristic polynomial of $1 +\gamma +Q$
which must be real. It is interesting that this puts restrictions on the
possible values of $Q$: given a fixed value of $\gamma$, for large enough
$Q$ some roots will be complex. To put it differently, from our proof that
these theories are conformally invariant we infer that if a matrix
$XAX^{-1}$ is diagonal for a real matrix $A$ and a real, symmetric,
invertible matrix $X$, then all the roots of the characteristic polynomial
of $A$ are real.

This unfortunately means that the large-$Q$ scenario of
\cite{Fortin:2011bm}, which leads to interesting oscillatory behavior in
unparticle physics, is excluded by conformal invariance. More generally,
the constraints that unitarity and scale invariance alone place on the
scaling dimensions of operators are weaker than those that follow if
conformal invariance is also imposed \cite{Grinstein:2008qk}. These weaker
conditions are popular in unparticle phenomenology as they amplify the
putative effects of unparticles. Of course our proof does not rule out
theories that are scale-invariant but not conformal outside the realm of
perturbation theory, leaving a smidgen of hope for unparticle enthusiasts.

\newsec{The \texorpdfstring{$\mathbf{\emph{c}}$}{c}-theorem in the presence
of cycles}[Sec:a-theorem]
As we have seen, the consistency relations of JO lead to the $c$-theorem in
perturbation theory,
\eqn{-\frac{d\widetilde B_b}{dt}=
\tfrac18\chi_{IJ}^g B_I B_J\ge0,}[strongc]
with $\widetilde B_b$ defined in \Btildeb.  Only the last step in this
sequence of relations invokes perturbation theory, for the positivity of
the metric $\chi_{ij}^g$ is established perturbatively.   For a
non-perturbative proof we turn to the method of KS.

Let us review the argument of KS.  Our presentation is closer in spirit to
that of LPR. We will try to note explicitly when implicit assumptions in
that argument are made. While plausible, these assumptions should be
justified for the theorem to be established. We deviate from both
presentations in that we do not derive nor use a Wess--Zumino dilaton
action for, as we will see, this is not necessary for the computation.

Consider the four point function of the operator
$\tfrac12\partial_\mu(x_\nu T^{\mu\nu})$ in an arbitrary four-dimensional
theory which is classically scale-invariant.  Furthermore, we will consider
kinematics such that $p_i^2=0$, $i=1,\ldots,4$, for the momenta $p_i$ of
the four insertions, so that the Mandelstam variables satisfy $s + t + u =
0$. Equivalently, for the theory on a conformally flat background,
$g_{\mu\nu}=e^{-2\tau(x)}\eta_{\mu\nu}$, one may compute the $\tau(x)$
scattering amplitude $A(s,t)$ with the on-shell condition $\nabla^2\tau=0$.

Now, we will assume that the forward scattering amplitude
$\mathcal{A}_\text{fwd}(s)=\mathcal A(s,0)$ exists, that is that the limit
$t\to0$ of $\mathcal A(s,t)$ exists. This could fail if $\mathcal{A}(s,t)$
had terms of the form, e.g., $\sim s^2\ln(t)$. Now,
$\mathcal{A}_\text{fwd}(s)$ can be computed by taking four
$\tau(x)$-derivatives of the generating functional and then taking the
metric as flat, the coupling constants to be spacetime-independent and the
background field $A_\mu$ and the conformal compensator to vanish.
Alternatively, and more straightforwardly, one can work with a conformally
flat metric  and having the only spacetime dependence in couplings and
$A_\mu$ arise through the dependence on the conformal factor $\tau(x)$, so
that one merely needs to take $\tau(x)=0$ after four times differentiating
$\widetilde W$. Now the first derivative simply gives the conformal anomaly
equation
\eqn{T^\mu_{\phantom{\mu}\!\mu}
={\widehat B}_I[\CO_I]+[\beta^{\mathscr{Q}}]+[(D^\mu\phi)^T B_\mu^A \phi]
-\mu^{-\epsilon}\beta_{\tilde\lambda}\cdot\mathscr{R}
+\nabla_\mu (J_\Theta^\mu+\widetilde{\mathscr{Z}}^\mu)
-((1+\hat\gamma+S)\phi)\cdot\frac{\delta}{\delta\phi}\tilde S_0,}
One need only thrice differentiate this equation to obtain the four-point
amplitude of $T^\mu_{\phantom{\mu}\!\mu}$. Note that on fixed points and
cycles, where we will need this, the first term vanishes since $\widehat
B_I=0$.  Also, the last term, which vanishes by the equations of motion,
can be ignored for the computation of the amplitude. Most of the remaining
terms vanish once the couplings are taken to be spacetime-independent (and
the metric flat and $A_\mu=0$). The remaining terms arise from the $G$ and
$H^2$ terms in $\beta_\lambda\cdot\mathscr{R}$. For a conformally flat
metric, $\gamma_{\mu\nu}=\exp(-2\tau(x))\eta_{\mu\nu}$, one has (in $d$
spacetime dimensions)
\eqna{e^{-4\tau}G&=8(\partial^2\tau)^2
-8\tau_{,\mu\nu}\tau^{,\mu\nu}-16\tau_{,\mu}\tau_{,\nu}\tau^{,\mu\nu}
- 8(d-3)\tau_{,\mu}\tau^{,\mu}\partial^2\tau
+2(d-1)(d-4)(\tau_{,\mu}\tau^{,\mu})^2\\
e^{-4\tau}H^2&=4(\partial^2\tau)^2
- 4(d-2)\tau_{,\mu}\tau^{,\mu}\partial^2\tau
+(d-2)^2(\tau_{,\mu}\tau^{,\mu})^2.}
The cubic term in $H^2$ vanishes for an ``on-shell'' conformal factor
$\partial^2\tau=0$ and so the only contribution to the ``on-shell'' forward
scattering amplitude is from $G$:
\eqn{\mathcal{A}_\text{fwd}(s)|_{\text{FP or cycle}}
=-\beta_b(s^2+t^2+u^2)|_{t=0} =-2\beta_b s^2.}[AFWDlimit]
Let's assume that there exists an RG trajectory from a UV fixed point or
cycle to an IR fixed point or cycle. On this trajectory this equation no
longer holds. However, we can inspect limiting behavior. Since
$\mathcal{A}_\text{fwd}/s^2$ depends on $s$ only through the dimensionless
ratio $\mu^2/s$, its behavior is dictated by the renormalization group.
Hence,
\eqn{\lim_{s\to\infty}\frac{\mathcal{A}_\text{fwd}(s)}{s^2}
=\lim_{s\to\infty}\frac{\mathcal{A}_\text{fwd}(s)|_{\text{FP or
cycle}}}{s^2}=-2(\beta_b)_{\text{UV}}}
and
\eqn{\lim_{s\to0}\frac{\mathcal{A}_\text{fwd}(s)}{s^2}
=\lim_{s\to0}\frac{\mathcal{A}_\text{fwd}(s)|_{\text{FP or
cycle}}}{s^2}=-2(\beta_b)_{\text{IR}}, }
where $(\beta_b)_{\text{UV}}$ and $(\beta_b)_{\text{IR}}$ are the limiting
UV and IR values of $\beta_b$ along the trajectory and correspond to those
on the fixed point or cycle. LPR study the approach to these limiting
values using conformal perturbation theory.

\begin{figure}[ht]
  \centering
  \begin{tikzpicture}
    \draw[->] (0,-1)--(0,4.5);
    \draw[->] (-4.2,0)--(4.2,0);
    \begin{scope}[>=latex',thick,line cap=round,line join=round]
      \draw[->-] (-3.8,0.2)--(-0.2,0.2) node[midway,above] {$I_2\;\;\;$};
      \draw[->-] (0.2,0.2)--(3.8,0.2) node[midway,above] {$I_2\;\;\;$};
      \draw (0.2,0.2) arc (0:180:0.2);
      \draw[->-] (3.8,0.2) arc (0:90:3.8);
      \draw[-<-] (-3.8,0.2) arc (180:90:3.8);
      \node at (-0.8,1.25) {$I_1$};
      \node at (3.05,3.05) {$I_3$};
      \node at (-3.05,3.05) {$I_3$};
    \end{scope}
    \draw[->] (-0.7,1) .. controls (-0.2,0.2) and (-0.7,1.4) .. (-0.1,0.38);
    \draw (3.5,4.5)--(3.5,4.1)--(3.9,4.1) node[above left=-0.5pt] {$s$};
  \end{tikzpicture}
  \caption{The contour of integration for
  $\displaystyle\int\frac{ds}{s^3}\,\mathcal{A}_\text{fwd}(s)$.}
  \label{fig:contour}
\end{figure}
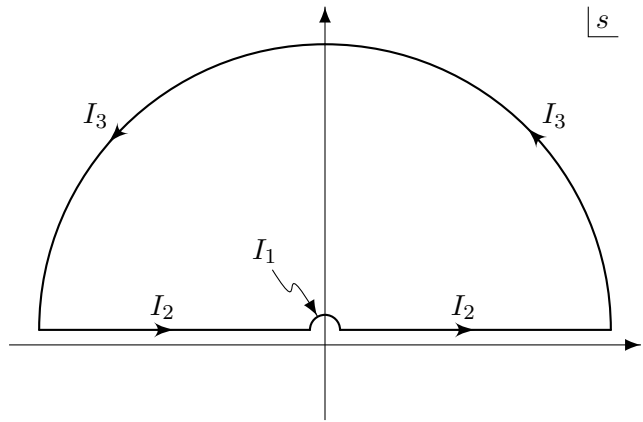

Following LPR we next consider the integral of
$\mathcal{A}_\text{fwd}(s)/s^3$ over the contour in Fig.~\ref{fig:contour}.
The integral over the semicircle $I_1$ cannot
be easily computed, but in the limit that the radius of the semicircle
vanishes it is reasonable that one can use the limiting value,
\eqn{ \int_{I_1}\frac{ds}{s^3} \mathcal{A}_\text{fwd}(s)\approx
 \int_{I_1}\frac{ds}{s}2(\beta_b)_\text{IR}=2\pi i (\beta_b)_\text{IR},}[IRlimit]
where the last step corresponds to taking the vanishing limit of the radius
of the semicircle $I_1$. Similarly, the large circle $I_3$ gives
\eqn{\int_{I_3}\frac{ds}{s^3} \mathcal{A}_\text{fwd}(s)\approx
\int_{I_3}\frac{ds}{s}2(\beta_b)_\text{UV}=-2\pi i (\beta_b)_\text{UV}.}[UVlimit]
It follows from Cauchy's theorem that
\eqna{ (\beta_b)_\text{UV}-(\beta_b)_\text{IR}
&=\frac1{2\pi i}\int_{I_2}
\frac{ds}{s^3} \mathcal{A}_\text{fwd}(s)\\
&=\frac1{\pi }\int_{0}^\infty \frac{ds}{s^3}
\Ima(\mathcal{A}_\text{fwd}(s+i0)),}
where in the last line LPR assume crossing symmetry to write
$\mathcal{A}_\text{fwd}(-s+i0)=\mathcal{A}^*_\text{fwd}(s+i0)$. Finally,
the KS argument invokes the optical theorem that relates the imaginary part
of the forward scattering amplitude to a positive-definite cross section to
conclude that
\eqn{(\beta_b)_\text{UV}-(\beta_b)_\text{IR}>0.}
We note in passing that the optical theorem is known to apply for forward
scattering amplitudes of (on-shell) physical particles. It is not clear a
priori that it applies to Green functions of composite operators at
$p_i^2=0$, even if it corresponds to the scattering amplitude of would-be
dilaton scattering. We think the assumption of positivity is reasonable,
so we press on.

What steps in the argument above require special attention when the theory
admits dimension-three currents? As we have pointed out, the trace of the
stress-energy tensor now has an additional $\partial_\mu J^\mu$ term, but
we have already accounted for this in the presentation above: the current
can be eliminated by replacing $B_I$ for $\beta_I$ in the expression for
the trace of the stress-energy tensor. Throughout the flow this makes no
difference to the argument above, since the positivity of the integral over
the segments $I_2$ of the contour follows from the optical theorem.  For
cycles one is not free to ignore the $\tau(x)$ dependence of the couplings
or the background vector field in the anomaly equation. But on the cycle
the couplings are covariantly constant.  Hence, the terms that vanish at
fixed points because of the constancy of couplings also vanish for cycles,
but now because they are covariantly constant. Finally, the validity of the
limits in \IRlimit and \UVlimit needs to be established anew for limit
cycles. However, the same method of conformal perturbation theory may be
applied to establish the result.  Since it is only scaling that is used in
this step of the argument by LPR, the proof goes through as presented
there.

\newsec{Summary and concluding remarks}[Sec:conclusions]
We have shown that the Komargodski--Schwimmer proof of the weak version of
the $c$-theorem includes the more general case that a renormalization group
flow goes from a fixed point or cycle to another fixed point or cycle.
Regarding the strong version of the $c$-theorem, proven in perturbation
theory by Jack and Osborn, we pointed out that the quantity that plays the
role of $c$ is $\widetilde B_b$ (defined in \Btildeb) which is closely
related to the $a$-anomaly ($\beta_b$ in the notation of Jack and Osborn);
these quantities agree at fixed points and on cycles, but are not generally
the same.

We presented a calculation of the Lie-algebra function of coupling
constants $S$ introduced by Jack and Osborn. This is the first calculation
of $S$ to an order (third) in the loop expansion where it does not vanish.
We then proved that $S=0$ on fixed points and that $S$ precisely
corresponds to the generator $Q$ of limit cycles when evaluated at any
point on the limit cycle.  This gives a  major improvement on the method of
searching for limit cycles: one merely needs to find zeroes of the beta
functions to the first order in the loop expansion (second order for
Yang--Mills couplings) and evaluate $S$ there. If $S=0$ the zero
corresponds to a fixed point, while if $S= Q\ne0$ the zero corresponds to a
limit cycle with $Q$ the generator of the cycle.

We used these results to show that the trace of the stress-energy tensor
vanishes on cycles, and hence that scale implies conformal invariance
(perturbatively in unitary relativistic $d=4$ QFT). If ``theory space'' is
understood as the space of couplings of a model modulo the action of $G_F$
on these couplings (with $G_F$ the group of symmetries of the free
Lagrangian), then cycles and fixed points are mapped to single points.  It
is remarkable that all such points describe in fact CFTs.

Some questions remain which we intend to turn to in the future.  Among them
are:
\begin{itemize}
\item Are there renormalization group flows between fixed points and
  cycles?
\item Are there limit cycles in four dimensions with bounded tree-level
  scalar potential?
\item Can a non-perturbative proof of the strong version of the
  $c$-theorem be given by extending the perturbative proof, say,
  by showing positivity of the metric $\chi^g_{IJ}$ using
  dispersion relations?
\item Do relativistic, unitary QFTs admit recursive RG flows that
  do not correspond to motions by generators in $G_F$?
\end{itemize}
We look forward to addressing these questions.

\ack{We are happy to acknowledge discussions with Hugh Osborn, Markus Luty,
Joseph Polchinski and Riccardo Rattazzi that challenged our previous
conclusion that theories on RG limit cycles are only scale-invariant.  We
thank David Pirtskhalava for numerous discussions and for his comments on
the manuscript, as well as Ken Intriligator for discussions at the
initiation of this project. This material is based upon work supported by
the US Department of Energy under contract DOE-FG03-97ER40546 and in part
by the US National Science Foundation under Grant No.\ 1066293 and the
hospitality of the Aspen Center for Physics.}

\begin{appendices}
\newsec{Ambiguities in RG functions}[app:Bfunction]
It is well-known that anomalous-dimension matrices and beta functions are
dependent on the renormalization scheme.  Nevertheless, physical quantities
obtained from the anomalous-dimension matrices and the beta functions which
are relevant to the study of scale-invariant theories are, as expected,
independent of the scheme \cite{Fortin:2012ic}.

It is however usually not appreciated that anomalous-dimension matrices and
beta functions exhibit another freedom, mentioned briefly in the beginning
of Section \ref{Sec:DimThree}, which we review here. For simplicity
consider a theory of real multi-component scalars with bare Lagrangian
\eqn{\mathscr{L}_0=\tfrac12\partial_\mu\phi_{0a}\partial^\mu\phi_{0a}
-\tfrac{1}{4!}g^0_{abcd}\phi_{0a}\phi_{0b}\phi_{0c}\phi_{0d}.}
There is an ambiguity in the definition of the wavefunction renormalization
matrix $Z^{1/2}$, corresponding to the freedom of choosing
$\tilde{Z}^{1/2}=OZ^{1/2}$ where $O^TO=\mathds{1}$ \cite{Jack:1990eb}. In
this appendix we study the effect of this ambiguity in the definition of RG
functions. For simplicity we present this analysis in the flat background
limit. Dimensional regularization is used throughout.

Bare couplings and fields are related to the corresponding renormalized
quantities by
\eqn{g_{I}^0=\mu^{k_I\epsilon}(g_I+L_I(g)),\qquad\phi_0=
\mu^{\delta\epsilon}\check Z(g)\phi,}
where $\check Z=Z^{1/2}$, and $\check Z-1$ and $L_I$ have expansions in
$\epsilon$-poles starting at $1/\epsilon$. The anomalous-dimension matrices
and the beta functions, as well as the antisymmetric matrix $S$ of \Sdef,
are given by
\eqn{\hat\gamma=\delta\epsilon-k_Ig_I\partial_I\check Z^1,\qquad
\hat\beta_I=-k_Ig_I\epsilon-k_IL_I^1+k_Jg_J\partial_JL_I^1,\qquad
S=-k_Ig_IN_I^1,}
where the superscript denotes residues of simple poles. The index carried
by $k$ is exempt from the summation convention. In the present example
$k_I=1$, but we keep it for generality.  Since we are interested in
ambiguities that arise because of different choices in the subtraction of
infinite quantities, we assume that $O$ has an expansion in
$\epsilon$-poles, $O=1+O^1/\epsilon+\cdots$, where $O^1$ is antisymmetric
as required by $O^TO=1$.  Then, under the freedom mentioned above, it is
easy to verify that the relevant quantities change as
\eqn{\check Z^1\to \check Z^1+O^1,\qquad
L_I^1\to L_I^1+(O^1g)_I,\qquad
N_I^1\to N_I^1-\partial_IO^1,}
This induces a change in the anomalous-dimension matrix, the beta
functions, and the antisymmetric matrix $S$:
\eqn{\hat\gamma\to\hat\gamma-\omega,\qquad
\hat\beta_I\to\hat\beta_I+(\omega g)_I,\qquad
S\to S+\omega,}
where $\omega=k_Ig_I\partial_IO^1$. This ambiguity, or ``gauge'' freedom,
in the definition of anomalous dimensions and beta functions is usually
resolved by requiring that the anomalous-dimension matrix be symmetric.
Note, however, that the trace of the stress-energy tensor, being a physical
quantity, has to be invariant under this unphysical freedom. Indeed, this
is obviously the case in \traceT. As we see $\hat\beta$, $\hat\gamma$ and
$S$ are gauge-covariant, but $\widehat B_I=\hat\beta_I-(Sg)_I$ and
$\widehat\Gamma= \hat\gamma+S$ are gauge-invariant. Although RG flows are
specified by $\hat\beta$, there is a gauge, defined by $\omega=-S$ so that
$S=0$, in which $\widehat B=\hat\beta$ and $\widehat\Gamma = \hat\gamma$.

Finally, it is worth pointing out that $\widehat B_I$ could be seen as the
proper vector field whose RG flows one should consider, and whose fixed
points describe CFTs. This vector field does not admit cycles in
perturbation theory.

\newsec{The relation between the metrics
\texorpdfstring{$\chi^a_{ij}$}{chi a} and
\texorpdfstring{$\chi^g_{ij}$}{chi g}}[app:chigvschia]
The coefficient $c_g$ of LPR appears to play the role of the ``metric''
$\chi^g_{ij}$ in the consistency condition \JOWeylconditionA.  As we
mention in the end of Section \ref{sec:WZ} and elaborate on further here,
this is not the case. To see the connection with the work of JO,
following LPR we write
\eqn{\Delta W_{\text{anomaly}}=\frac12\int d^4x\, d^4y\,\tau(x)\tau(y)
\vev{\Theta(x)\Theta(y)},}
where $\Theta=\beta^i\CO_i$,  and therefore
\eqn{\frac{d}{dt}\Delta W_{\text{anomaly}}=\frac12 \int d^4x\,d^4y\,\tau(x)\tau(y)
\frac{d}{dt}\vev{\Theta(x)\Theta(y)}.}[dDeltaWdt]
In Ref.~\cite[Eq.~(3.18b)]{Osborn:1991gm} Osborn finds the RGE for the
product of two local renormalized operators,
\eqn{-\frac{d}{dt}\vev{\CO_i(x)\CO_j(0)} + \partial_i\beta^j
\vev{\CO_i(x)\CO_j(0)} + \partial_j\beta^k\vev{\CO_i(x)\CO_j(0)}=
-\chi_{ij}^a\partial^2\partial^2\delta^{(4)}(x).}
The quantity $\chi_{ij}^a$ can be thought of as the beta function
associated with the counterterm needed in order to renormalize the
correlator $\vev{\CO_i(x)\CO_j(0)}$. Now since
$-d\beta^i/dt=\beta^j\partial_j\beta^i$, it is easy to see that
\eqn{\frac{d}{dt}\vev{\Theta(x)\Theta(0)}=
\chi_{ij}^a\beta^i\beta^j\partial^2\partial^2\delta^{(4)}(x).}[RunTheta]
Using this in \dDeltaWdt we see that the metric of LPR is $-2\chi^a$, which
is always positive. This suggests the question ``is there a relation
between $\chi^g$ and $\chi^a$?''

In the specific example of a gauge theory with a simple gauge group $G$ and
charged Dirac fermions in some representation, JO
give~\cite[Eqs.~(5.12)]{Jack:1990eb}, at two loops,
\eqn{\chi^{a(2)}=-\frac12 \chi^{g(2)}=-\frac{n_V}{8\pi^2 g^2}\left[
1+\left(17C-\frac{20}{3}R\right)h \right],\qquad
h\equiv\frac{g^2}{16\pi^2},}[ChiaChigTwoL]
where $\tr(t^a_{\text{adj}}t^b_{\text{adj}})=C\delta^{ab}$, $R$ is
similarly defined for the representation of the Dirac fermions, and
$n_V=\dim(\text{Adj})$ is the number of vectors.  However, the relation
$\chi^g=-2\chi^a$ of \ChiaChigTwoL does not hold in general, and so the
task of computing $\chi^g$ is complicated.  Nevertheless, Weyl consistency
conditions give the general relation between $\chi^a$ and
$\chi^g$~\cite[Eq.~(3.23)]{Jack:1990eb}:
\eqn{\chi^g_{ij}+2\chi_{ij}^a-\bar{\chi}_{ijk}^a\beta^k = -\beta^k
\partial_k V_{ij}-\partial_i\beta^j\,V_{kj}-\partial_j\beta^k\,
V_{ik},}[ChiaChigGen]
where $\zeta V_{ij}=\bar{\chi}_{ijk}^ak^kg^k$ (no sum over the index $k$),
and $\bar{\chi}^a_{ijk}=\partial_k\chi^a_{ij}-\tfrac12(\chi^b_{ikj}
+\chi^b_{jki})$, with $\chi^b_{ijk}$ necessary to regulate infinities in
three-point functions, and $\zeta$ defined as an operator counting the
number of loops, whose form can be read off from $\CO(\epsilon)$ terms of
the finiteness condition (3.9e) of JO:
\eqn{\zeta V_{ij}=(1+k^k g^k\partial_k)V_{ij}+2k^iV_{ij} \qquad
(\text{no sum over the index carried by }k)}
(cf.\ JO's (3.16b)).

In our gauge-theory example \ChiaChigGen becomes
\eqn{\chi^g+2\chi^a-\bar{\chi}^a\beta^g=-\beta^g\frac{\partial V}{\partial
g}-2\frac{\partial\beta^g}{\partial g}V,\qquad \zeta V=\tfrac12
\bar{\chi}^a g,}[ChiaChig]
where $\zeta V=(2+\frac12 g\,\partial/\partial g)V=(2+h\,\partial/\partial
h)V$, the beta function for the gauge coupling is
\eqn{\frac{1}{g}\beta^g=-\beta_0h -\beta_1 h^2+ \CO(h^3) ,\qquad
\beta_0=\frac13(11C-4R),\qquad \beta_1=\frac23C(17C-10R),}
and $\bar{\chi}^a=\partial\chi^a/\partial g-\chi^b$, where $\chi^b$ is
given at two loops by $\chi^{b(2)}=\frac{n_V}{4\pi^2g^3}(1+4\beta_0h)$.
It follows that
\eqn{\bar{\chi}^a=-\frac{n_V}{\pi^2g^3}[\beta_0h+\CO(h^2)].}
Expanding $V=v_0+v_1h+\cdots$ gives $\zeta V= 2v_0+3v_1h+\cdots=\frac12
g\bar\chi^a$, or
\eqn{V=-\frac{n_V\beta_0}{64\pi^4}+\CO(h).}
With these results \ChiaChig gives
\eqn{\chi^g+2\chi^a=-\frac{n_V\beta_0^2}{32\pi^4}h+\CO(h^2),
}[ChiaChigGauge]
and, therefore, beyond two-loop order, $\chi^g\neq -2\chi^a$.

To summarize, the results of LPR correspond to using JO's $-2\chi_{ij}^a$
as a metric, which however is not in general equal to JO's metric
$\chi_{ij}^g$. Indeed, $\chi^g_{ij}+2\chi^a_{ij}$ fails to vanish beyond
the first few orders in the loop expansion. The positivity of $\chi^g_{ij}$
may also fail non-perturbatively (for example, if its perturbative
expression has finite radius of convergence).

\newsec{How to calculate \texorpdfstring{$N_I$}{N\_I} and
\texorpdfstring{$S$}{S}}[app:S]
The calculation of JO's $N_I$ proceeds order by order in perturbation
theory. In this appendix we calculate contributions to $N_I$ in a quantum
field theory with real scalars and Weyl spinors up to two loops, and we
also perform a three-loop calculation of the part of $N_I$ that is needed
in order to compute $S$.

As can be seen from \prenewct, in order to calculate $N_I$ we need to
compute self-energies of scalars but with coupling constants as spectator
fields. Equivalently, the calculation can be done by considering scalar
self-energy diagrams and letting momentum come in from external legs and go
out through couplings.  From these diagrams we can then pick up the
contribution linear in the momentum of the field and linear in the momentum
of the coupling. After we antisymmetrize, we have a contribution to $N_I$.

It is perhaps helpful to remind the reader here that in a theory with
scalars and fermions the $I$ index can be either $(abcd)$ or $(a|ij)$. Let
us also remark that $S$ appears first at three loops in a theory with
scalars and spinors. The reason is easily seen from \Sdef: a diagram that
contributes to $N$ will only contribute to $S$ if it is not symmetric under
$a\leftrightarrow b$. As it turns out there are no such diagrams in scalar
self-energies at one and two loops, but there are four such diagrams at
three loops. Consequently, even if the theory contains gauge fields,
diagrams with gauge fields will not contribute to $S$ at three loops, but
certainly will do so at higher order.  Therefore, even in a gauge theory we
don't need to include gauge fields in our leading-order calculation of $S$.

\begin{figure}[H]
  \centering
  \begin{tikzpicture}
    \begin{scope}
      \draw[dashed] (0,0)--(1.5,0)
                    (5,0)--(3.5,0);
      \draw[->] (0.45,0.2)--(1.05,0.2) node[midway,above] {$p$};
      \filldraw[black] (1.5,0) circle (1pt);
      \filldraw[black] (3.5,0) circle (1pt);
      \draw[->] (3.5,0)--+(45:0.6cm) node[right] {$p$};
      \draw (2.5,0) circle (1cm);
    \end{scope}
    \begin{scope}[xshift=6cm]
      \draw[dashed] (0,0)--(1.5,0)
                    (5,0)--(3.5,0);
      \draw[->] (4.55,0.2)--(3.95,0.2) node[midway,above] {$\;p$};
      \filldraw[black] (1.5,0) circle (1pt);
      \filldraw[black] (3.5,0) circle (1pt);
      \draw[->] (1.5,0)--+(135:0.6cm) node[left] {$p$};
      \draw (2.5,0) circle (1cm);
    \end{scope}
      \begin{scope}[xshift=0.5cm,yshift=-2.5cm]
      \draw[dashed] (0,0)--(2,0)
                    (4,0)--(2,0);
      \draw[->] (0.7,0.2)--(1.3,0.2) node[midway,above] {$p$};
      \draw[fill=white] (2,0) circle (5pt);
      \draw[->] (2,0)--+(45:0.7cm) node[right] {$p$};
      \draw (2,0)--+(45:5pt)
            (2,0)--+(225:5pt)
            (2,0)--+(135:5pt)
            (2,0)--+(315:5pt);
    \end{scope}
    \begin{scope}[xshift=6.5cm,yshift=-2.5cm]
      \draw[dashed] (0,0)--(2,0)
                    (4,0)--(2,0);
      \draw[->] (3.3,0.2)--(2.7,0.2) node[midway,above] {$\;p$};
      \draw[fill=white] (2,0) circle (5pt);
      \draw[->] (2,0)--+(135:0.7cm) node[left] {$p$};
      \draw (2,0)--+(45:5pt)
            (2,0)--+(225:5pt)
            (2,0)--+(135:5pt)
            (2,0)--+(315:5pt);
    \end{scope}
  \end{tikzpicture}
  \caption{Diagrams that contribute to $N_{a|ij}$ at one loop and their
  corresponding counterterms.}\label{fig:NOneLoop}
\end{figure}
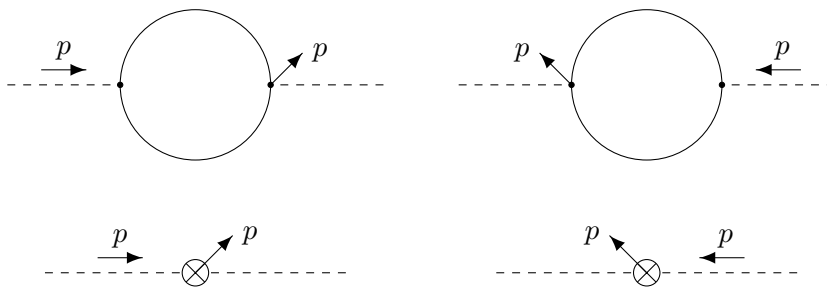

\subsec{One loop}
At one loop the calculation proceeds with no subtleties since
renormalization is trivial, i.e., there are no subdivergences to be
subtracted.  The two diagrams that contribute to $N_I$ and their
corresponding counterterms are shown in Fig.~\ref{fig:NOneLoop}.
\noindent A straightforward calculation gives
\eqn{(N_{c|ij})_{ab}=-\frac{1}{16\pi^2\epsilon}\,\frac{1}{2}
(y^\ast_{a|ij}\delta_{bc}-y^\ast_{b|ij}\delta_{ac})+\text{finite},}
and there is of course a complex conjugate $(N^\ast_{c|ij})_{ab}$.

In order to simplify the notation we write the result for the residue of
the simple $\epsilon$-pole in $N_I$ in the form
\eqn{16\pi^2 (N^1_I)_{ab}\partial^\mu g_I=-\tfrac{1}{2}[\tr(y_a^{\phantom{\ast}}\partial^\mu y_b^\ast)
+\text{h.c.}-\{a\leftrightarrow b\}],}
where $g_I$ on the left-hand side stands here for $y_{c|ij}$ or
$y_{c|ij}^\ast$. Selecting the appropriate derivatives one easily reads off
the corresponding $N^1_I$. Our result reproduces JO's equation (7.16) for
$\rho_I$ when we use Dirac spinors.

\begin{figure}[ht]
  \centering
  \begin{tikzpicture}
    \begin{scope}
      \draw[dashed] (0,0)--(1.5,0)
                    (5,0)--(3.5,0)
                    (1.5,0)--(3.5,0);
      \draw[dashed,dash phase=-1.2pt] (3.5,0) arc [start angle=0,end
      angle=180,radius=1cm];
      \draw[dashed,dash phase=-1.2pt] (1.5,0) arc [start angle=180,end
      angle=360,radius=1cm];
      \filldraw[fill=black] (1.5,0) circle (1pt);
      \filldraw[fill=black] (3.5,0) circle (1pt);
    \end{scope}
    \begin{scope}[xshift=5.5cm]
      \draw[dashed] (0,0)--(1.5,0)
                    (5,0)--(3.5,0)
                    (2.5,1)--(2.5,-1);
      \draw (2.5,0) circle (1cm);
      \filldraw[fill=black] (1.5,0) circle (1pt)
                            (3.5,0) circle (1pt)
                            (2.5,1) circle (1pt)
                            (2.5,-1) circle (1pt);
    \end{scope}
    \begin{scope}[xshift=11cm]
      \draw[dashed] (0,0)--(1.5,0)
                    (5,0)--(3.5,0);
      \node[name=s,circle,draw,minimum height=2cm,xshift=2.5cm] {};
      \draw[dashed,dash phase=-1.8pt] (s.north west) .. controls (2,0.2)
      and (3,0.2) .. (s.north east);
      \filldraw[fill=black] (1.5,0) circle (1pt)
                            (3.5,0) circle (1pt)
                            (s.north east) circle (1pt)
                            (s.north west) circle (1pt);
    \end{scope}
  \end{tikzpicture}
  \caption{Feynman diagrams that contribute to $N_I$ at two
  loops.}\label{fig:NTwoLoops}
\end{figure}
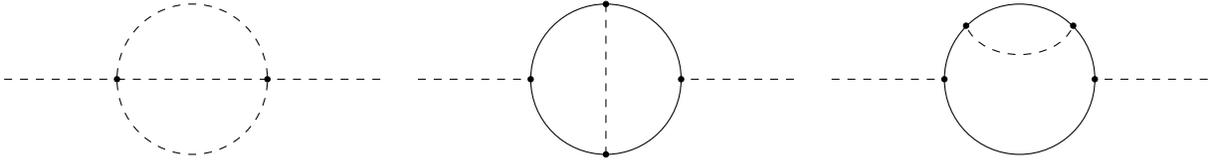

\subsec{Two loops}
At two loops there are three Feynman diagrams that contribute to $N_I$,
listed in Fig.~\ref{fig:NTwoLoops}.
The calculation of the residues of the simple $\epsilon$-poles of $N_I$
requires now a subtraction of subdivergences, something that proceeds, for
the most part, in the usual way. However, there is a small subtlety, not
seen in the usual treatments of renormalization, that we would like to
point out. Clearly, the two right-most diagrams of Fig.~\ref{fig:NTwoLoops}
have subdivergences so we have to add to them the diagrams with the
insertions of the corresponding counterterms. For the right-most diagram
the graph with the insertion of the counterterm is
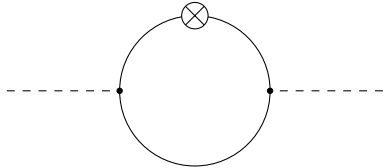
\begin{figure}[H]
  \centering
  \begin{tikzpicture}
    \draw[dashed] (0,0)--(1.5,0)
                  (5,0)--(3.5,0);
    \node[circle,draw,minimum height=2cm,xshift=2.5cm] {};
    \filldraw[cross,fill=white] (2.5,1) circle (5pt);
    \filldraw[fill=black] (1.5,0) circle (1pt)
                          (3.5,0) circle (1pt);
  \end{tikzpicture}
\end{figure}
\noindent Now, when the momentum that comes in from, say the left external
leg, flows out through the counterterm, then there are two diagrams that
contribute, namely
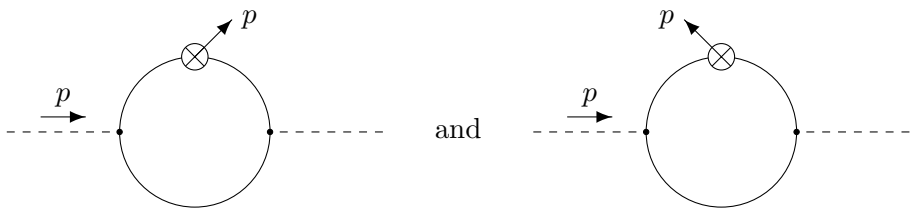
\begin{figure}[H]
  \centering
  \begin{tikzpicture}
    \begin{scope}
      \draw[dashed] (0,0)--(1.5,0)
                    (5,0)--(3.5,0);
      \draw[->] (0.45,0.2)--(1.05,0.2) node[midway,above] {$p$};
      \node[circle,draw,minimum height=2cm,xshift=2.5cm] {};
      \filldraw[cross,fill=white] (2.5,1) circle (5pt);
      \draw[->] (2.5,1)--+(45:0.7cm) node[right] {$p$};

      \filldraw[fill=black] (1.5,0) circle (1pt)
                            (3.5,0) circle (1pt);
    \end{scope}
    \begin{scope}[xshift=6cm]
      \node at (0,0.05) {and};
    \end{scope}
    \begin{scope}[xshift=7cm]
      \draw[dashed] (0,0)--(1.5,0)
                    (5,0)--(3.5,0);
      \draw[->] (0.45,0.2)--(1.05,0.2) node[midway,above] {$p$};
      \node[circle,draw,minimum height=2cm,xshift=2.5cm] {};
      \draw[->] (2.5,1)--+(135:0.7cm) node[left] {$p$};
      \filldraw[cross,fill=white] (2.5,1) circle (5pt);
      \filldraw[fill=black] (1.5,0) circle (1pt)
                            (3.5,0) circle (1pt);
    \end{scope}
  \end{tikzpicture}
\end{figure}
\noindent where the momentum exits to the north-east or to the north-west
depending on which vertex it flows out of in the original diagram in
Fig.~\ref{fig:NTwoLoops}. In both cases the counterterm is the same, but
the diagram with the insertion of the counterterm is different as a result
of the difference in the momentum of the internal leg that the counterterm
picks up. That is, had we retained different momenta for the various
vertices, there would be two momenta associated with the  counterterm.

The two-loop result for $N^1_I$, previously unpublished,  is
\eqna{(16\pi^2)^2(N^1_I)_{ab}\partial^\mu
g_I=&-\tfrac{1}{24}\lambda_{acde}\,\partial^\mu\lambda_{bcde}+[\tfrac{1}{4}\tr(y_a^\ast y_c^{\phantom{\ast}}\partial^\mu y_b^\ast y_c^{\phantom{\ast}})\\
&+\tfrac{1}{8}\tr(y_a^\ast\partial^\mu y_c^{\phantom{\ast}}y_c^\ast y_b^{\phantom{\ast}})+\tfrac{3}{8}\tr(y_a^\ast y_c^{\phantom{\ast}}y_c^\ast\partial^\mu y_b^{\phantom{\ast}})+{\rm h.c.}]-\{a\leftrightarrow b\}.}
It follows that $S$ vanishes at this order. This can be seen, term by term
(when anti-symmetrized in $a$ and $b$) by replacing $g_I$ for $\partial^\mu
g_I$.

\begin{figure}[ht]
  \centering
  \begin{tikzpicture}
    \begin{scope}
      \draw[dashed] (0,0)--(1.5,0)
                    (5,0)--(3.5,0)
                    (3.5,0)--(2.5,0);
    \draw (2.5,1)--(2.5,-1);
    \draw (2.5,1) arc [start angle=90, end angle=270, radius=1cm];
    \draw[dashed] (2.5,-1) arc [start angle=270, end angle=360, radius=1cm];
    \draw[dashed] (3.5,0) arc [start angle=0, end angle=90, radius=1cm];
    \foreach \position in {(1.5,0), (2.5,0), (2.5,1), (2.5,-1), (3.5,0)}{
      \filldraw[fill=black] \position circle (1pt);}
    \end{scope}
    \begin{scope}[xshift=6cm,rotate around={180:(2.5,0)}]
      \draw[dashed] (0,0)--(1.5,0)
                    (5,0)--(3.5,0)
                    (3.5,0)--(2.5,0);
    \draw (2.5,1)--(2.5,-1);
    \draw (2.5,1) arc [start angle=90, end angle=270, radius=1cm];
    \draw[dashed] (2.5,-1) arc [start angle=270, end angle=360, radius=1cm];
    \draw[dashed] (3.5,0) arc [start angle=0, end angle=90, radius=1cm];
    \foreach \position in {(1.5,0), (2.5,0), (2.5,1), (2.5,-1), (3.5,0)}{
      \filldraw[fill=black] \position circle (1pt);}
    \end{scope}
    \begin{scope}[yshift=-3cm]
      \draw[dashed] (0,0)--(1.5,0)
                    (5,0)--(3.5,0)
                    (2.5,1)--(2.5,-1);
      \draw (1.5,0) arc [start angle=180,end angle=150,radius=1cm]
      coordinate (left point);
      \draw (2.5,1) arc [start angle=90,end angle=120,radius=1cm]
      coordinate (right point);
      \draw let \p1=(left point), \p2=(right point) in
      (\x1,\y1) to [controls=+(120:0.3) and +(150:0.3)] (\x2,\y2);
      \draw[dashed,dash phase=0.1pt] let \p1=(left point), \p2=(right
      point) in (\x1,\y1) to [controls=+(330:0.3) and +(300:0.3)]
      (\x2,\y2);
      \draw (2.5,1) arc [start angle=90,end angle=0,radius=1cm];
      \draw (1.5,0) arc [start angle=180,end angle=360,radius=1cm];
      \foreach \position in {(1.5,0), (2.5,1), (2.5,-1),(3.5,0)}{
        \filldraw[fill=black] \position circle (1pt);}
      \filldraw[fill=black] let \p1=(left point) in (\x1,\y1) circle
      (1pt);
      \filldraw[fill=black] let \p2=(right point) in (\x2,\y2) circle
      (1pt);
    \end{scope}
    \begin{scope}[xshift=6cm,yshift=-3cm,rotate around={180:(2.5,0)}]
      \draw[dashed] (0,0)--(1.5,0)
                    (5,0)--(3.5,0)
                    (2.5,1)--(2.5,-1);
      \draw (1.5,0) arc [start angle=180,end angle=150,radius=1cm]
      coordinate (left point);
      \draw (2.5,1) arc [start angle=90,end angle=120,radius=1cm]
      coordinate (right point);
      \draw let \p1=(left point), \p2=(right point) in
      (\x1,\y1) to [controls=+(120:0.3) and +(150:0.3)] (\x2,\y2);
      \draw[dashed,dash phase=0.1pt] let \p1=(left point), \p2=(right
      point) in (\x1,\y1) to [controls=+(330:0.3) and +(300:0.3)]
      (\x2,\y2);
      \draw (2.5,1) arc [start angle=90,end angle=0,radius=1cm];
      \draw (1.5,0) arc [start angle=180,end angle=360,radius=1cm];
      \foreach \position in {(1.5,0), (2.5,1), (2.5,-1),(3.5,0)}{
        \filldraw[fill=black] \position circle (1pt);}
      \filldraw[fill=black] let \p1=(left point) in (\x1,\y1) circle
      (1pt);
      \filldraw[fill=black] let \p2=(right point) in (\x2,\y2) circle
      (1pt);
    \end{scope}
  \end{tikzpicture}
  \caption{Three-loop diagrams that contribute to $N_I$ not symmetric under
    $a\leftrightarrow b$, and thus leading to contributions to $S$ at three
    loops.}\label{fig:NSThreeLoops}
\end{figure}
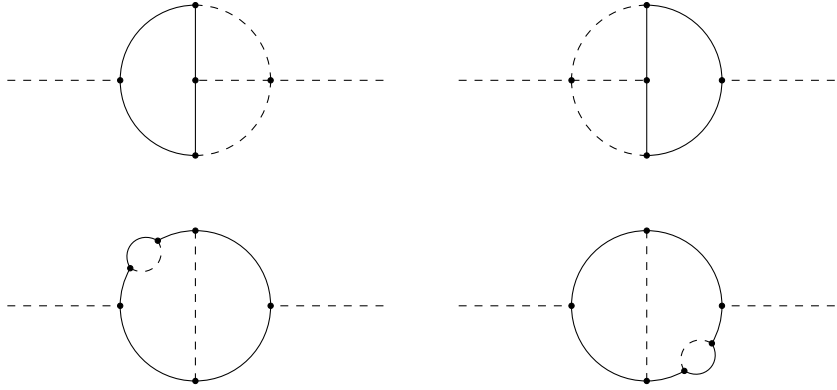

\subsec{Three loops}
At three loops there are many diagrams that contribute to $N_I$, but only
four are not symmetric under $a\leftrightarrow b$ and thus end up
contributing to $S$. These diagrams are shown in
Fig.~\ref{fig:NSThreeLoops}, and we here only compute their contributions
to $N_I^1$.
From these diagrams (and the corresponding counterterms), using the methods
for the calculation of pole parts of three-loop diagrams given in
\cite{Chetyrkin:1997fm}, we find
\eqna{(16\pi^2)^3(N^1_I)_{ab}\partial^\mu g_I\supset&
-\tfrac{1}{2}\tr(y^{\phantom{\ast}}_a\partial^\mu
y^\ast_cy^{\phantom{\ast}}_dy^\ast_e)\lambda_{bcde}
-\tfrac{1}{3}\tr(y^{\phantom{\ast}}_ay^\ast_c\partial^\mu
y^{\phantom{\ast}}_dy^\ast_e)\lambda_{bcde}
-\tfrac{1}{2}\tr(y^{\phantom{\ast}}_ay^\ast_cy^{\phantom{\ast}}_d\partial^\mu
y^\ast_e)\lambda_{bcde}\\
&-\tfrac{5}{24}\tr(y^{\phantom{\ast}}_ay^\ast_cy^{\phantom{\ast}}_dy^\ast_e)
\partial^\mu\lambda_{bcde}
-\tfrac{1}{24}\tr(y^{\phantom{\ast}}_b\partial^\mu
y^\ast_cy^{\phantom{\ast}}_dy^\ast_e)\lambda_{acde}
-\tfrac{5}{24}\tr(y^{\phantom{\ast}}_by^\ast_c\partial^\mu
y^{\phantom{\ast}}_dy^\ast_e)\lambda_{acde}\\
&-\tfrac{1}{24}\tr(y^{\phantom{\ast}}_by^\ast_cy^{\phantom{\ast}}_d\partial^\mu
y^\ast_e)\lambda_{acde}
-\tfrac{5}{24}\tr(\partial^\mu
y^{\phantom{\ast}}_by^\ast_cy^{\phantom{\ast}}_dy^\ast_e)\lambda_{acde}
-\tfrac{7}{32}\tr(y^{\phantom{\ast}}_a\partial^\mu
y^\ast_cy^{\phantom{\ast}}_dy^\ast_dy^{\phantom{\ast}}_by^\ast_c)\\
&-\tfrac{7}{96}\tr(y^{\phantom{\ast}}_ay^\ast_c\partial^\mu
y^{\phantom{\ast}}_dy^\ast_dy^{\phantom{\ast}}_by^\ast_c)
-\tfrac{23}{96}\tr(y^{\phantom{\ast}}_ay^\ast_cy^{\phantom{\ast}}_d\partial^\mu
y^\ast_dy^{\phantom{\ast}}_by^\ast_c)
-\tfrac{7}{96}\tr(y^{\phantom{\ast}}_ay^\ast_cy^{\phantom{\ast}}_dy^\ast_d\partial^\mu
y^{\phantom{\ast}}_by^\ast_c)\\
&-\tfrac{7}{32}\tr(y^{\phantom{\ast}}_ay^\ast_cy^{\phantom{\ast}}_dy^\ast_d
y^{\phantom{\ast}}_b\partial^\mu y^\ast_c)
+\tfrac{1}{16}\tr(y^{\phantom{\ast}}_a\partial^\mu
y^\ast_cy^{\phantom{\ast}}_cy^\ast_dy^{\phantom{\ast}}_by^\ast_d)
-\tfrac{5}{48}\tr(y^{\phantom{\ast}}_ay^\ast_c\partial^\mu
y^{\phantom{\ast}}_cy^\ast_dy^{\phantom{\ast}}_by^\ast_d)\\
&-\tfrac{1}{48}\tr(y^{\phantom{\ast}}_ay^\ast_cy^{\phantom{\ast}}_c \partial^\mu
y^\ast_dy^{\phantom{\ast}}_by^\ast_d)
-\tfrac{7}{96}\tr(y^{\phantom{\ast}}_ay^\ast_cy^{\phantom{\ast}}_cy^\ast_d \partial^\mu
y^{\phantom{\ast}}_by^\ast_d)
+\tfrac{1}{16}\tr(y^{\phantom{\ast}}_ay^\ast_cy^{\phantom{\ast}}_cy^\ast_d
y^{\phantom{\ast}}_b\partial^\mu y^\ast_d)\\
&+{\rm h.c.}-\{a\leftrightarrow b\},}
and since
\eqn{S\equiv-k_IN^1_Ig_I=-N^1_{abcd}\lambda_{abcd}-(\tfrac12 N^1_{a|ij}y_{a|ij}+\text{h.c.})}
we finally obtain
\eqn{(16\pi^2)^3S_{ab}=\tfrac{5}{8}\tr(y^{\phantom{\ast}}_ay^\ast_c
y^{\phantom{\ast}}_dy^\ast_e)\lambda_{bcde}
+\tfrac{3}{8}\tr(y^{\phantom{\ast}}_ay^\ast_cy^{\phantom{\ast}}_d
y^\ast_dy^{\phantom{\ast}}_by^\ast_c)+{\rm h.c.}-\{a\leftrightarrow b\}.}
As already remarked in the main body, evaluating this on points in coupling
space where we have found fixed points and cycles in
Refs.~\cite{Fortin:2011ks, Fortin:2012ic,Fortin:2012cq}, we find that $S$
vanishes at all fixed points and equals $Q$ on all cycles.

\end{appendices}
\bibliography{Cyclic_CFTs_ref}

\end{document}